\documentclass[sigconf,screen]{acmart}

\AtBeginDocument{%
  \providecommand\BibTeX{{%
    \normalfont B\kern-0.5em{\scshape i\kern-0.25em b}\kern-0.8em\TeX}}}

\setcopyright{acmlicensed}
\acmPrice{15.00}
\acmDOI{10.1145/3597926.3598063}
\acmYear{2023}
\copyrightyear{2023}
\acmSubmissionID{issta23main-p96-p}
\acmISBN{979-8-4007-0221-1/23/07}
\acmConference[ISSTA '23]{Proceedings of the 32nd ACM SIGSOFT International Symposium on Software Testing and Analysis}{July 17--21, 2023}{Seattle, WA, USA}
\acmBooktitle{Proceedings of the 32nd ACM SIGSOFT International Symposium on Software Testing and Analysis (ISSTA '23), July 17--21, 2023, Seattle, WA, USA}
\received{2023-02-16}
\received[accepted]{2023-05-03}

\newcommand{\para}[1]{\vspace{2pt}\noindent\textbf{#1.~}}
\newcommand{\ignore}[1]{}

\usepackage{ulem}
\usepackage[switch]{lineno}
            
\usepackage{listings, xcolor}
\definecolor{verylightgray}{rgb}{.97,.97,.97}
\lstdefinelanguage{Solidity}{
  keywords=[1]{anonymous, assembly, assert, balance, break, call, callcode, case, catch, class, constant, continue, constructor, contract, debugger, default, delegatecall, delete, do, else, emit, event, experimental, export, external, false, finally, for, function, gas, if, implements, import, in, indexed, instanceof, interface, internal, is, length, library, log0, log1, log2, log3, log4, memory, modifier, new, payable, pragma, private, protected, public, pure, push, require, return, returns, revert, selfdestruct, send, solidity, storage, struct, suicide, super, switch, then, this, throw, transfer, true, try, typeof, using, value, view, while, with, addmod, ecrecover, keccak256, mulmod, ripemd160, sha256, sha3}, 
  keywordstyle=[1]\color{blue}\bfseries,
  keywords=[2]{address, bool, byte, bytes, bytes1, bytes2, bytes3, bytes4, bytes5, bytes6, bytes7, bytes8, bytes9, bytes10, bytes11, bytes12, bytes13, bytes14, bytes15, bytes16, bytes17, bytes18, bytes19, bytes20, bytes21, bytes22, bytes23, bytes24, bytes25, bytes26, bytes27, bytes28, bytes29, bytes30, bytes31, bytes32, enum, int, int8, int16, int24, int32, int40, int48, int56, int64, int72, int80, int88, int96, int104, int112, int120, int128, int136, int144, int152, int160, int168, int176, int184, int192, int200, int208, int216, int224, int232, int240, int248, int256, mapping, string, uint, uint8, uint16, uint24, uint32, uint40, uint48, uint56, uint64, uint72, uint80, uint88, uint96, uint104, uint112, uint120, uint128, uint136, uint144, uint152, uint160, uint168, uint176, uint184, uint192, uint200, uint208, uint216, uint224, uint232, uint240, uint248, uint256, var, void, ether, finney, szabo, wei, days, hours, minutes, seconds, weeks, years},  
  keywordstyle=[2]\color{teal}\bfseries,
  keywords=[3]{block, blockhash, coinbase, difficulty, gaslimit, number, timestamp, msg, data, gas, sender, sig, value, now, tx, gasprice, origin},  
  keywordstyle=[3]\color{violet}\bfseries,
  identifierstyle=\color{black},
  sensitive=false,
  comment=[l]{//},
  morecomment=[s]{/*}{*/},
  commentstyle=\color{gray}\ttfamily,
  stringstyle=\color{red}\ttfamily,
  morestring=[b]',
  morestring=[b]"
}
\usepackage{pifont}
\usepackage{diagbox}
\usepackage{subfigure}

\usepackage{color}
\usepackage{microtype}

\usepackage{oplotsymbl}
\usepackage{subfigure}
\usepackage{siunitx}
\usepackage{array,framed}
 \usepackage{
   color,
   float,
   epsfig,
   wrapfig,
   graphics,
   graphicx,
 }
\usepackage{setspace}
\usepackage{amsfonts}
\usepackage{latexsym,fancyhdr,url}
\usepackage{enumerate}
\usepackage{algorithm2e}
\usepackage{algpseudocode}
\usepackage{graphics}
\usepackage{xparse} 
\usepackage{xspace}
\usepackage{multirow}
\usepackage{appendix}
\usepackage{microtype}

\usepackage{soul}
\soulregister\cite7 
\soulregister\ref7 

\usepackage{color}

\usepackage{fancyhdr}

\usepackage{subfigure}
\begin{document}

\title{Definition and Detection of Defects in NFT Smart Contracts}

\author{Shuo Yang}
\affiliation{%
  \institution{Sun Yat-sen University}
  \city{Zhuhai}
  \country{China}
}
\email{yangsh233@mail2.sysu.edu.cn}

\author{Jiachi Chen}
\authornote{corresponding author}
\affiliation{%
  \institution{Sun Yat-sen University}
  \city{Zhuhai}
  \country{China}
}
\email{chenjch86@mail.sysu.edu.cn}

\author{Zibin Zheng}
\affiliation{%
  \institution{Sun Yat-sen University}
  \city{Zhuhai}
  \country{China}}
\email{zhzibin@mail.sysu.edu.cn}

\renewcommand{\shortauthors}{Shuo Yang, Jiachi Chen, and Zibin Zheng}

\begin{abstract}
Recently, the birth of non-fungible tokens (NFTs) has attracted great attention. NFTs are capable of representing users' ownership on the blockchain and have experienced tremendous market sales due to their popularity. Unfortunately, the high value of NFTs also makes them a target for attackers. The defects in NFT smart contracts could be exploited by attackers to harm the security and reliability of the NFT ecosystem. Despite the significance of this issue, there is a lack of systematic work that focuses on analyzing NFT smart contracts, which may raise worries about the security of users' NFTs. To address this gap, in this paper, we introduce 5 defects in NFT smart contracts. 
Each defect is defined and illustrated with a code example highlighting its features and consequences, paired with possible solutions to fix it.
Furthermore, we propose a tool named NFTGuard to detect our defined defects based on a symbolic execution framework. Specifically, NFTGuard extracts the information of the state variables from the contract abstract syntax tree (AST), which is critical for identifying variable-loading and storing operations during symbolic execution. Furthermore, NFTGuard recovers source-code-level features from the bytecode to effectively locate defects and report them based on predefined detection patterns. We run NFTGuard on 16,527 real-world smart contracts and perform an evaluation based on the manually labeled results. We find that 1,331 contracts contain at least one of the 5 defects, and the overall precision achieved by our tool is 92.6\%.
\end{abstract}

\ccsdesc[500]{Software and its engineering~Software verification and validation}


\keywords{smart contracts, NFTs, defects definition and detection, symbolic execution}


\maketitle

\vspace{-0.12cm}
\section{Introduction}
\label{sec:intro}
A non-fungible token (NFT) is a representation of the ownership of a digital or physical asset on the blockchain~\cite{das2021understanding}. Not only digital pictures, but also videos and physical artworks such as paintings can be turned into NFTs for trading. The popularity of NFTs has attracted many people to join. In 2021, the trading volume of NFTs reached approximately 41 billion USD in cryptocurrency~\cite{siakam_nft_2022}. Smart contracts are Turing-complete programs that run on the blockchain~\cite{zheng2017overview}, which enable diverse decentralized applications (DApps) to deploy on it~\cite{zheng2018blockchain}.
As the underlying technology of NFT projects, smart contracts enable developers to encode rules in NFT trading and allow users to mint and transfer NFTs in the trading markets. The ERC-721 standard~\cite{eip721} is proposed by the Ethereum Improvement Proposals (EIPs) to allow for the implementation of a standard API for NFTs within smart contracts. The standard provides basic functionality to track and transfer NFTs, enabling NFT smart contracts to operate effectively~\cite{wang2021non}.

The high value of NFTs makes the security of NFT contracts especially important. In addition, due to the immutability of smart contracts, it is critical to ensure that the NFT smart contract is bug-free before it is deployed on the blockchain. A contract defect is an error, flaw, or fault that may cause unexpected results or change the original intentions of the code~\cite{lyu1996handbook}. Contract defects are not only related to security issues but also design flaws which may make the contract risky in the future; thus, defective NFT smart contracts can significantly harm the NFT ecosystem and cause a heavy loss of users. For example, an attacker exploited a reentrancy defect to drain approximately 1,300 ETH (1.43 million USD) from the NFT money market platform Omni~\cite{omni_hacker}. The Akutar NFT lost 34 million USD due to a design flaw~\cite{blocksec_how_2022}. Detecting defects in smart contracts could help us to find and remove bugs, thus improving the security of the contract and avoiding severe consequences.

Although a set of smart contract defects have been reported in previous work~\cite{chen2020defining}, many scenarios cannot be covered due to the increasing complexity and security requirements of smart contracts, e.g., NFT smart contracts. To address the problem, we first conducted an empirical study to define smart contract defects by analyzing collected 672 Q\&A posts on StackOverflow~\cite{stackoverflow}, and 88 security analysis reports that expose security issues in NFT smart contracts released by some blockchain security companies. Based on the manual filtering and open card sorting approach~\cite{spencer2009card}, we then filtered and analyzed 31 StackOverflow posts and 29 security analysis reports. Finally, we define 5 defects in NFT smart contracts from different security scenarios: \textit{Risky Mutable Proxy}, \textit{ERC-721 Reentrancy}, \textit{Unlimited Minting}, \textit{Missing Requirements} and \textit{Public Burn}. For each defect, we propose possible solutions for improving the quality and robustness of the programs.

To expose the defective NFT smart contracts, we developed a tool named NFTGuard to detect the defined 5 defects in real-world smart contracts.
The design of NFTGuard combines the source code level information and bytecode to make detection more effective in complicated NFT contracts. 
Specifically, NFTGuard extracts state variable name, type, and storage information from the contract AST for identifying operations on state variables (e.g., variable reading and writing operations). NFTGuard recovers features on the source code level by identifying those operations, which helps us locate defects during symbolic execution at the bytecode level. NFTGuard reports the defects based on predefined patterns and a symbolic execution framework. We ran NFTGuard on 16,527 real-world NFT smart contracts and found that 1,331 of the smart contracts in our dataset contained at least one of the 5 defects. 
To evaluate the performance of NFTGuard, we manually labeled two datasets obtained through random sampling with a confidence interval of 10 and a confidence level of 95\%: one from contracts reported as defective; another from contracts with no reported defects. The results show that the overall precision of our tool is 92.6\%.

The main contributions of our work are as follows:
\vspace{-0.11cm}
\begin{itemize}
    \item We define 5 defects in NFT smart contracts from collected posts and security reports. We provide an illustration and a code example of each defect to help developers to better understand the defects. We also list the possible solutions for enhancing the security of the NFT ecosystem.
    \item We develop NFTGuard for detecting NFT smart contract defects based on a symbolic execution framework. NFTGuard recovers source code level features from bytecode for identifying designed defect patterns more effectively. NFTGuard is also an extensible framework with higher code coverage for analyzing smart contracts, which supports the latest Solidity compiler version, i.e., v0.8+.
    \item We run NFTGuard on 16,527 real-world smart contracts and evaluate its performance. We find that 1,331 of the 16,527 smart contracts contain at least one defined defect. Furthermore, the overall precision of our method is 92.6\% in our manually labeled dataset that we randomly selected from the reported defective contracts.
    \item We publicize the source code of NFTGuard, all of the experimental data, and analysis results with detailed markdown files at \href{https://github.com/NFTDefects/nftdefects}{https://github.com/NFTDefects/nftdefects}.
\end{itemize}
\vspace{-0.11cm}

This paper is organized as follows: Section~\ref{sec:background} offers background knowledge on smart contracts and the ERC-721 standard. Section~\ref{sec:defects} defines and provides examples of 5 defects. Section~\ref{sec:method} introduces NFTGuard's architecture and key techniques. Section~\ref{sec:evaluation} presents an experiment evaluating NFTGuard. Section~\ref{sec:discussion} discusses experimental results, tool design, and solutions for fixing defects. Section~\ref{sec:rw} reviews related works, while Section~\ref{sec:conclusion} provides the conclusion.

\vspace{-0.15cm}
\section{Background}
\label{sec:background}
\subsection{Solidity Smart Contract and EVM}\label{background:smart_contract}
A smart contract is ``a computerized transaction protocol that executes the terms of a contract''~\cite{szabo1997formalizing}. 
Solidity is the most popular programming language for smart contracts on Ethereum. When deploying a contract to Ethereum, the source code will be compiled into EVM bytecode and stored on the blockchain permanently. Due to the immutability of smart contracts, all deployed contracts are impossible to be modified even when bugs are detected.

The EVM is critical in running smart contracts deployed on Ethereum. It executes transactions by splitting contract EVM bytecode into operation codes (opcodes) and following their instructions.
However, EVM bytecode's jump positions can not be statically determined, and there are no return instructions for function calls~\cite{chen2019large}.
Data in smart contracts can be stored in storage, memory, or calldata~\cite{solidity}. Storage is for permanent data, while memory is for temporary use during contract execution. Each mutable state variable in a smart contract is assigned a slot ID during compilation, indicating its storage space. These slot IDs help identify the exact storage location of state variables during execution. For complex data types like mappings and dynamic arrays, locating storage areas requires hash calculations in collaboration with slot IDs~\cite{layout}.

\vspace{-0.1cm}
\subsection{ERC-721 Standard}
The ERC-721 standard, defined by EIPs, enables tracking of NFTs in smart contracts~\cite{eip721}. Tokens are digital assets on Ethereum's blockchain, which can be fungible or non-fungible. ERC-20 standard~\cite{eip20} applies to fungible, identical, and interchangeable tokens. In contrast, ERC-721-compliant NFTs are indivisible and unique, signifying distinct ownership of a specific digital or physical asset.

\vspace{-0.4cm}
\begin{figure}[htb]
\setlength{\abovecaptionskip}{0.cm}
\begin{lstlisting}[numbers=none]
function approve(address _approved, uint256 _tokenId) external payable;
function setApprovalForAll(address _operator, bool _approved) external;
function transferFrom(address _from, address _to, uint256 _tokenId) external payable;
function safeTransferFrom(address _from, address _to, uint256 _tokenId) external payable;
\end{lstlisting}
\caption{Important functions defined in ERC-721 standard.}
\label{fig:erc721}
\end{figure}
\vspace{-0.4cm}

The ERC-721 provides a standard API for NFT smart contracts and publicizes mandatory and optional interfaces. From the developer's perspective, the ERC-721 proposes development annotations for every interface, requiring developers to follow~\cite{eip721}.
For the mandatory interfaces, every ERC-721-compliant smart contract should implement the ERC-721 and ERC-165~\cite{eip165}. Figure~\ref{fig:erc721} lists some key functions related to our work. 
Specifically, the token owner can delegate their tokens to an operator to manage their assets through the \textit{setApprovalForAll} function. The \textit{approve} function can give the operation permission for a specific token to someone. The token owner, the approved operator, or a specific token operator can call the \textit{transferFrom} function to achieve the ownership transfer of a particular token. A ``safe'' function called \textit{safeTransferFrom} invokes function \textit{onERC721Received} claimed in the interface \textit{ERC721TokenReceiver}. A wallet or token receiver must implement \textit{onERC721Received} to support the token transfer. When a token flows to the contract, it will check whether the contract implements the interface. In addition, the token owner can set the token URI when minting a new token via the optional \textit{ERC721Metadata} extension. 
This interface enables users to query the details of the assets their NFTs represent.
Furthermore, the optional interface \textit{ERC721Enumerable} allows an NFT smart contract to publish its complete list of NFTs and make them discoverable.

In summary, the interfaces declared in the ERC-721 standard play an essential role in creating the NFT ecosystem. They mainly focus on permission control, e.g., ownership transfer, making NFTs capable of representing the ownership of users' assets on the blockchain.

\section{Defects in NFT Smart Contracts}
\label{sec:defects}
In this section, we illustrate how we obtain the 5 defects and provide a definition and example for each defect.
\vspace{-0.15cm}
\subsection{Data Collection}
\subsubsection{StackOverflow Posts}
To find and define contract defects for NFT smart contracts, we need to collect issues that developers have encountered and are concerned about. 
StackOverflow is a Q\&A platform that enables developers to communicate and share solutions for any problems they encounter. It provides powerful filtering functionality, such as tag filtering. A StackOverflow tag is a keyword or label that categorizes one question with other similar questions, and filtering by tags makes it easier to find related posts.
Therefore, we conducted the first round of filtering by using tag \textit{``NFT''} and \textit{``ERC721''}, and finally obtained 672 NFT-related StackOverflow posts for further analysis. 

\vspace{-0.15cm}

\subsubsection{Security Analysis Reports}
Security analysis reports are also ideal sources for categorizing defects in NFT smart contracts. In the past few years, NFTs have become increasingly popular, and their high value has attracted a large number of users. This has motivated some security teams to focus on the security of NFT-related smart contracts, producing many security analysis reports. We collected security analysis reports from a variety of sources, such as Medium~\cite{medium}, Twitter~\cite{twitter}, and the official websites of famous blockchain security teams such as SlowMist~\cite{slowmist} and PeckShield~\cite{peckshield}. Finally, we collected 88 reports related to NFT security and conducted further analysis.

\vspace{-0.15cm}
\subsection{Data Analysis}
\subsubsection{Manually Filtering}
In the previous subsection, we described the collection of 672 NFT-related StackOverflow posts and 88 security analysis reports. However, some of them might not be related to NFT smart contract defects. For example, although some posts are tagged with NFT, their contents are about web3js, e.g., how to interact with an NFT smart contract using web3js. In addition, some security reports revealed scams or phishing attacks about NFT projects unrelated to NFT contracts. Therefore, we manually removed the data irrelevant to smart contract defects.
Finally, after manual filtering, we found that a total of 31 out of 672 collected posts and 29 out of 88 collected security reports were related to Solidity-related NFT smart contract defects. 

\vspace{-0.15cm}
\subsubsection{Open Card Sorting}
To ensure the correctness of the results, we analyzed and categorized the filtered contract-defect-related posts and reports using the open card sorting~\cite{spencer2009card} approach. 
In this process, we consider two aspects to ensure the representative and significance of the defects, i.e., the reproducibility and popularity of the code issues. 
Some issues may be strongly related to a specific application and cannot be reproduced, e.g., a code issue in implementing a project's refunding business logic~\cite{blocksec_how_2022}. These issues are not classified as representative defects. 
Furthermore, we use view times and the amount of financial loss to analyze the popularity of issues in posts and security reports.

We created a card for each post and report, dividing the content into several parts. Two authors worked together to determine the labels for each post. We followed the detailed steps illustrated in~\cite{chen2020defining}. 
First, we randomly chose 40\% of the cards in our first classification round. 
We first read the title and description for StackOverflow posts to understand the defects the post was related to.
Then, we read the comments to understand how to deal with the defect.
We omitted the cards with no clear root cause and categorized the possible defects. Similarly, we read the description in the security analysis reports and checked the problematic code to find the root cause of the defect.

Next, in the second classification round, two authors independently categorized the remaining 60\% of the cards following the same steps described in the first round. 
After that, we compared our results and discussed the differences.
After discussion, we removed the defects that were uncommon and finally categorized the defects into 5 types. Among the classified posts and reports, the post views exceeded 41,000, and the financial loss exceeded 16 million USD.

\vspace{-0.4cm}
\begin{figure}[h]
    \setlength{\abovecaptionskip}{0.cm}
    \centering
    \includegraphics[width=\linewidth]{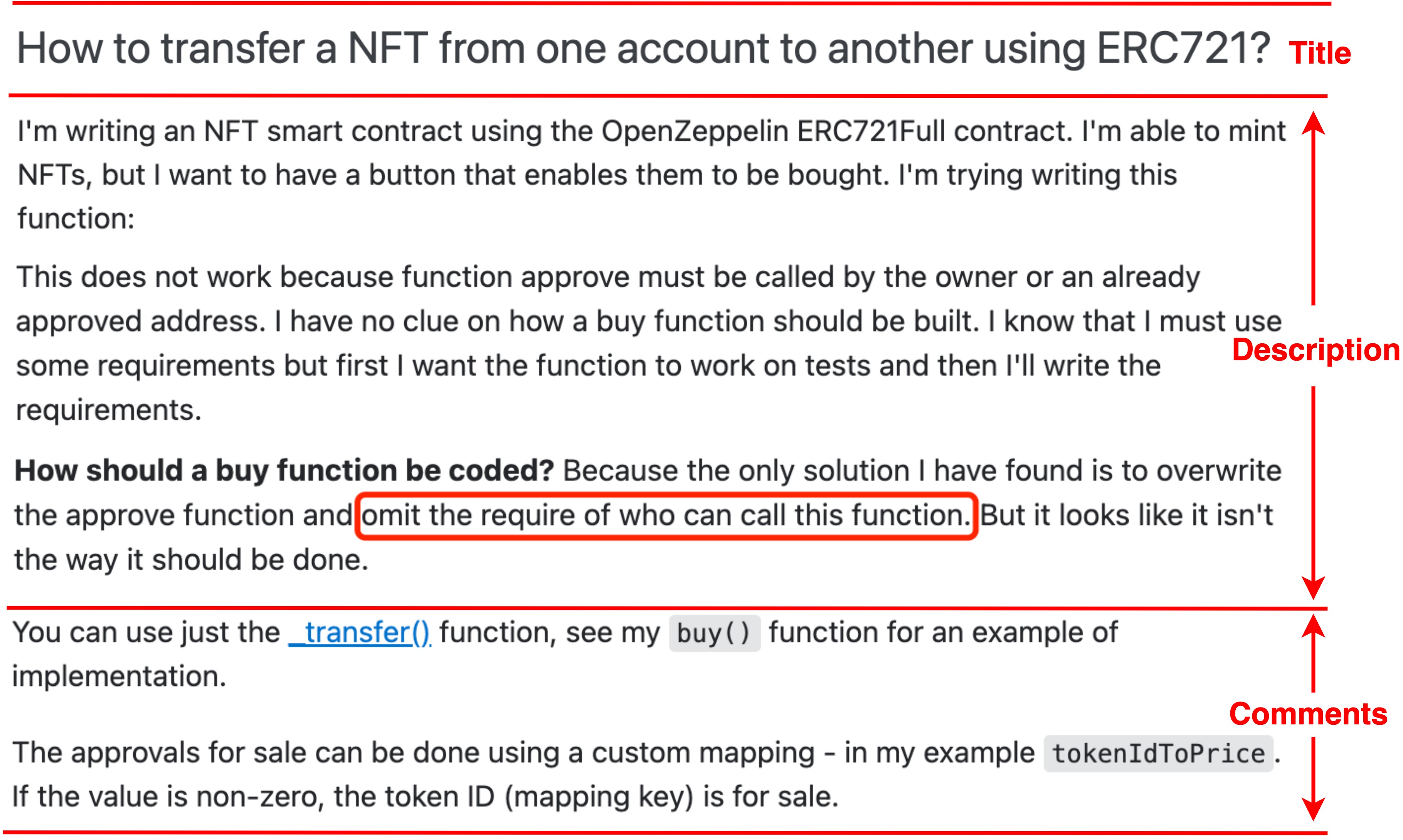}
    \caption{Example of a card of StackOverflow posts.} \label{fig:so}
\end{figure}
\vspace{-0.5cm}

Figure~\ref{fig:so} shows an example of a post reporting a defect. The card contains three parts, i.e., the title, description, and comments.
From the description of the card, it is not difficult to find that the developer seems unfamiliar with the ERC-721 standard and has trouble using the \textit{approve} function. Specifically, as shown in the red box, they intend to give approval to someone in a buy function but override the function \textit{approve} to skip the required check, which disobeys the standard. After that, we read the posted comments to determine the root cause of the error. Specifically, the first comment gives an optional solution to answer the question mentioned in the description. From the comment and corresponding question, we can determine that the root cause of the problem is the missing requirements to override the \textit{approve} function. Considering the reproducibility, we further discuss and generalize the problem to a common one related to missing requirements when implementing NFT smart contracts. Therefore, we obtain a category named \textit{``Missing Requirements''} from this post with over 16,000 views.

\begin{figure}[tb]
\setlength{\abovecaptionskip}{-0.1cm}
\setlength{\belowcaptionskip}{-0.4cm}
    \centering
    \includegraphics[width=3in]{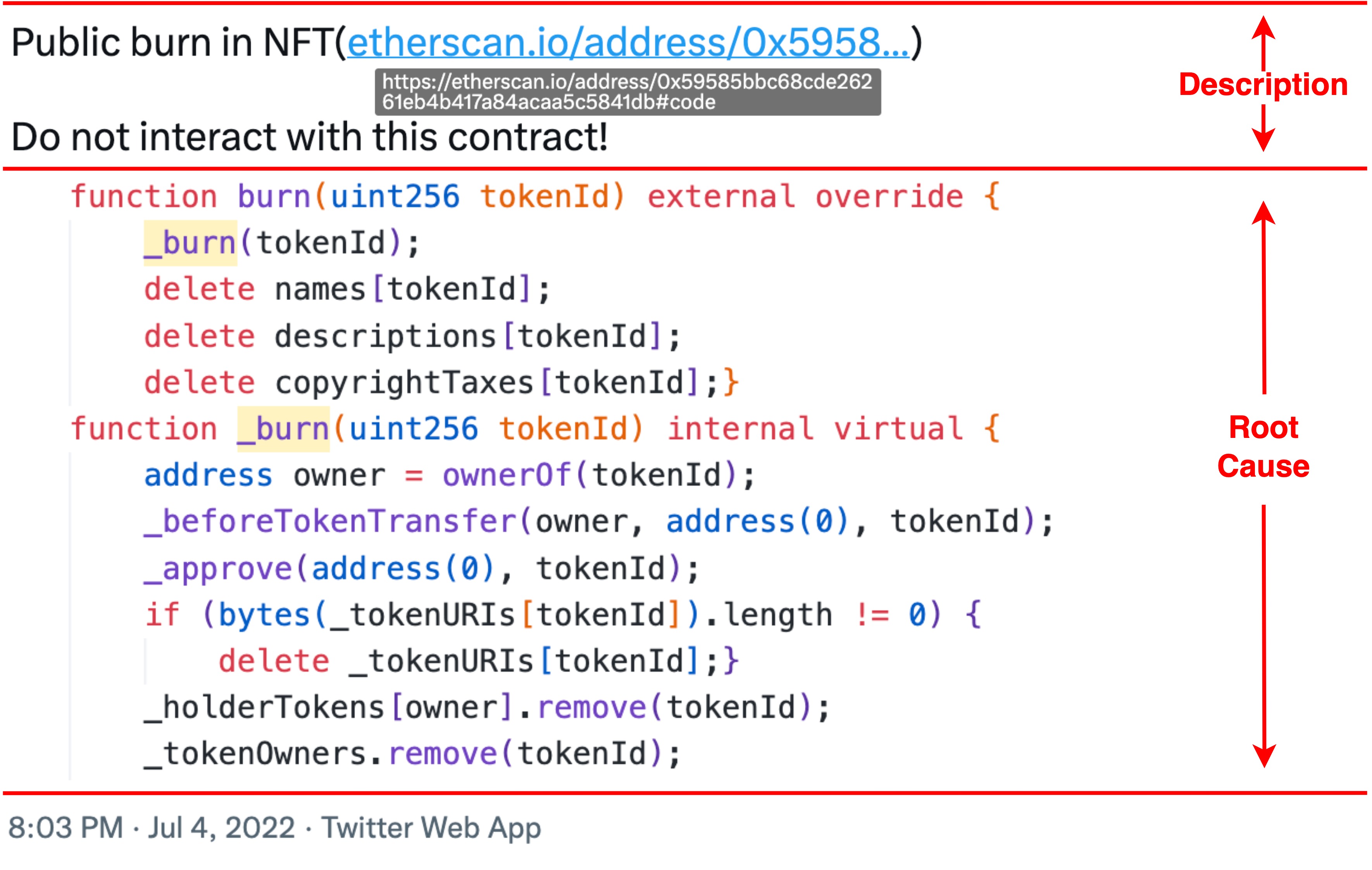}
    \caption{Example of a card of security analysis reports.} \label{fig:sar}
\end{figure}

Different from the StackOverflow posts, the security analysis reports reveal the root cause of a real-world security problem more directly. Figure~\ref{fig:sar} is an example of a security analysis report describing a defect. The card contains the problem description and the root cause of the problem. From the report, we can find that an NFT smart contract contains a public burn function. 
Then we try to find the problem mentioned in the card's root cause (e.g., real-world problematic code). We find that someone can burn others' NFTs without a permission check when executing the \textit{external} function \textit{burn}. The wrong permission setting is a reproducible problem; the financial loss was over 770K USD due to the problem. Therefore, we categorize a defect named \textit{``Public Burn''} from the card.

\vspace{-0.2cm}

\subsection{Defects Definition}
\label{sec:def}

Based on the methods introduced in the previous subsections, we finally define 5 defects related to NFT smart contracts. Attackers can utilize these defects to attack NFT projects and make unfair profits. We first give a brief definition of each NFT smart contract defect in Table~\ref{tab:defectdef}. Then, we provide the corresponding detailed definition and code example for each defect.

\vspace{-0.4cm}
\begin{table}[hb]
    \setlength\tabcolsep{3pt}
    \setlength{\abovecaptionskip}{0.cm}
    \begin{center}
        \caption{Definitions of the 5 Defects.}
        \label{tab:defectdef}
        \resizebox{\columnwidth}{!}{
        \begin{tabular}{l||p{5cm}}
            \hline
            \textbf{Contract Defect} & \textbf{Definition}  \\
            \hline
            \textit{Risky Mutable Proxy}               &  Make the proxy contract modifiable.                  \\
            \hline
            \textit{ERC-721 Reentrancy}        & Modify the state variable after the external invocation, which is determined by the variable.              \\
            \hline
            \textit{Unlimited Minting}         & Do not check the max supply of NFTs when minting.         \\
            \hline
            \textit{Missing Requirements}        &  Do not follow the annotations for developing ERC-721 standard interfaces.            \\
            \hline
            \textit{Public Burn}        & Do not check the caller of the burn operation on NFT.                    \\
             \hline
        \end{tabular}
        }
    \end{center}
\end{table}
\vspace{-0.4cm}

\textbf{(1) Risky Mutable Proxy:} 
OpenSea is the largest and the most popular NFT trading market~\cite{opensea} in the NFT ecosystem. It adopts the Wyvern protocol~\cite{wyvern} to facilitate the decentralized exchange of NFTs.
The first time a seller lists their NFTs in OpenSea, a proxy registry contract creates a smart contract called \textit{OwnableDelegateProxy}, which stores the seller's address. The proxy registry contract can use this new contract to take action on the seller's behalf and call methods on other contracts. When sellers list any item in their NFT smart contracts, they give their proxy registry contracts approval to transfer their tokens. Therefore, users do not need to pay extra gas for additional approvals for each NFT, making trading easy. However, since the proxy registry contract can control all of the user's tokens to perform trading for themselves, it is risky if the proxy registry address is modifiable. In that case, all of the user's tokens can be transferred to the attacker, who changes the proxy registry addresses through the proxy setting function without permission.

\textbf{Example:} In Figure~\ref{rp}, function \textit{setProxyRegistryAddress} (lines 1-2) enables external users, specifically the contract owner, to modify the proxy address. It is noticeable that even the contract owner can maliciously drain users' NFTs and make unfair profits. 
This procedure can be exploited to approve other addresses through the \textit{isApprovedForAll} function (lines 3-7), which returns true to allow other addresses to operate users' NFTs. The change in the proxy address can cause the ownership of NFTs to be transferred from their owners to attackers.

\vspace{-0.35cm}
\begin{figure}[htb]
\setlength{\abovecaptionskip}{0.cm}
\begin{lstlisting}
function setProxyRegistryAddress(address proxyAddress) external onlyOwner {
    proxyRegistryAddress = proxyAddress;}
function isApprovedForAll(address owner, address operator) override public view returns(bool){
    // Whitelist OpenSea proxy contract for easy trading.
    ProxyRegistry proxyRegistry = ProxyRegistry(proxyRegistryAddress);
    if (address(proxyRegistry.proxies(owner)) == operator) {return true;}
    return super.isApprovedForAll(owner, operator);}
\end{lstlisting}
\caption{An example of Risky Mutable Proxy defect.}
\label{rp}
\end{figure}
\vspace{-0.5cm}

\textbf{(2) ERC-721 Reentrancy:}
Different from the reentrancy pattern, which is related to the fallback function and ether transfer~\cite{liu2018reguard}, the invocation on a particular function causes the ERC-721 reentrancy, and no ether transfer happened in this process. The ERC-721 standard mandates the usage of the \textit{ERC721TokenReceiver} interface for receiving tokens. Noticeably, the token receiver can add other operations in the implemented \textit{onERC721Received} function claimed in \textit{ERC721TokenReceiver}. Therefore, someone can attack the victim contract by designing a malicious \textit{onERC721Received} function. For example, if a victim contract calls the \textit{safeTransferFrom} function that invokes the receiver's \textit{onERC721Received} function to check if the receiver is a contract, the malicious receiver can reenter the victim contract by adding a function callback. The intention of the contract can be interrupted while running. Most of the time, this reentrant call can cause a minted NFT to be more than the rarity threshold~\cite{das2021understanding}, hurting other buyers. Even worse, it can also cause severe financial loss~\cite{omni_hacker}.

\vspace{-0.4cm}
\begin{figure}[htb]
\setlength{\abovecaptionskip}{0.cm}
\begin{lstlisting}
function mintNFT(uint256 _numOfTokens, bytes memory _signature) public payable{
    /* precheck the addressMinted[msg.sender]=T/P */
    (bool success, string memory reason) = canMint(msg.sender, _signature);
    require(success, reason)
    for(uint i = 0; i < _numOfTokens; i++) {
        _safeMint(msg.sender, totalSupply() + 1);}
    addressMinted[msg.sender] = true;}
\end{lstlisting}
\caption{An example of ERC-721 Reentrancy defect.}
\label{reen}
\end{figure}
\vspace{-0.4cm}

\begin{figure}[htbp]
\setlength{\abovecaptionskip}{-0.1cm}
\setlength{\belowcaptionskip}{-0.5cm}
    \centering
    \includegraphics[width=2.8in]{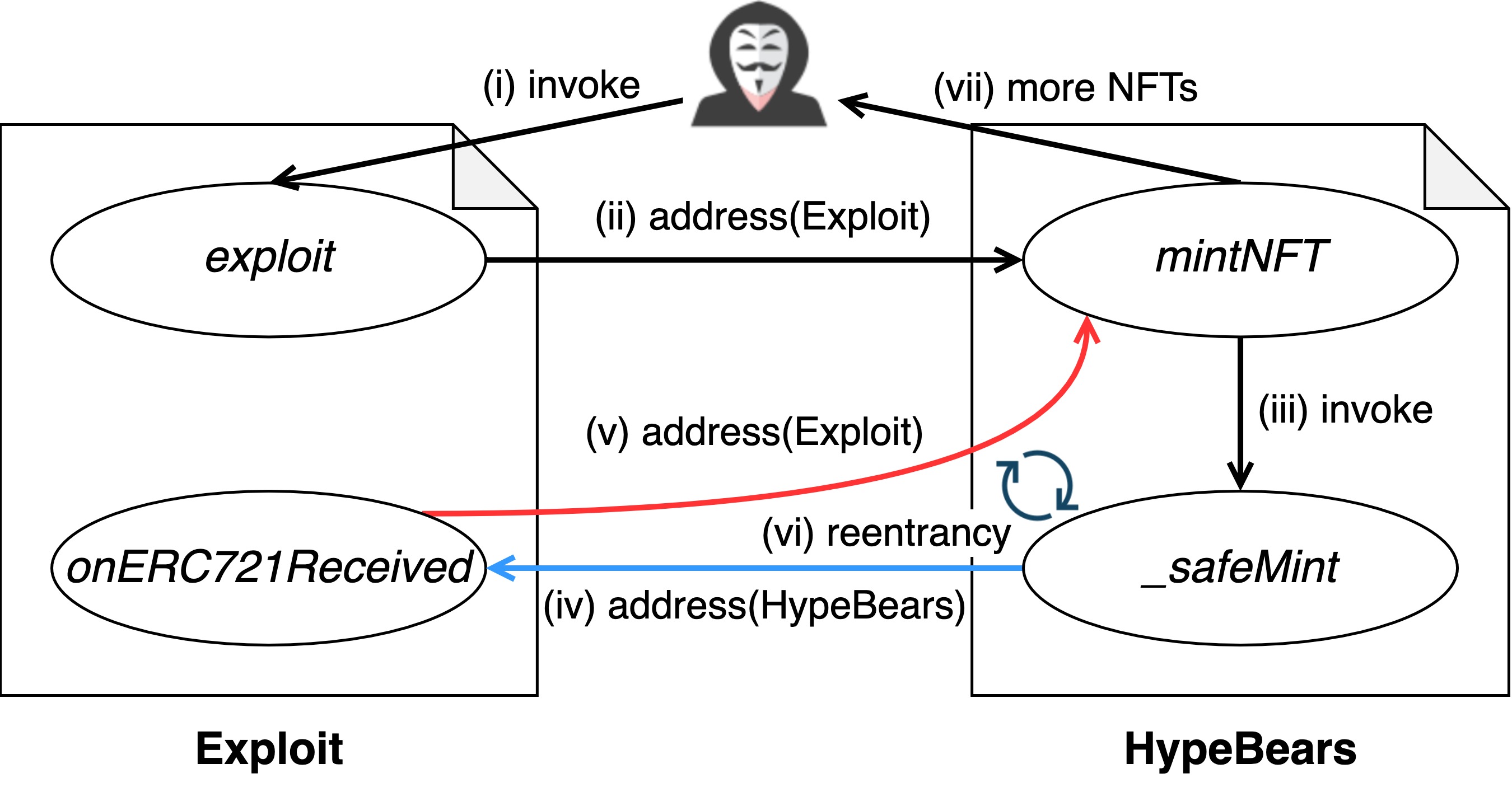}
    \caption{ERC-721 Reentrancy call to HypeBears.}
    \Description{An ERC-721 reentrancy sample.}
    \label{fig:hypebears}
\end{figure}

\textbf{Example:}
In Figure~\ref{reen}, function \textit{mintNFT} invokes the \textit{\_safeMint} function (line 6) of the OpenZeppelin implementation~\cite{erc721oz}. The \textit{\_safeMint} invokes \textit{IERC721Receiver.onERC721Received}, which can be utilized by the receiver (Contract \textit{Exploit} in Figure~\ref{fig:hypebears}) to perform a reentrant call. According to the contract inheritance and caller mechanism introduced in the official Solidity document~\cite{solidity}, the reentrant call is from the same address as the original caller to invoke \textit{mintNFT} (the red line in Figure~\ref{fig:hypebears}). As a result, the normal flow of the function can be interrupted. The status of \verb|addressMinted| (line 7) cannot be set to true, which means the caller can skip the check in the \textit{canMint} function (line 3) to mint more tokens than the threshold~\cite{blocksec_safemint}.

\textbf{(3) Unlimited Minting:}
A limited supply is one of the key features of the NFT ecosystem to maintain the digital scarcity~\cite{brekke_digital_2021} and control the abundance of a digital resource. 
Smart contracts can ensure scarcity by setting (1) the on-chain rarity parameter and (2) checking logic for prohibiting minting beyond promised limits, thereby imposing appropriate limits.
The \textit{totalSupply} function in \textit{ERC721Enumerable} can be used to obtain the number of minted NFTs to be compared with the max supply (rarity parameter) of the project. 
In many NFT projects, there is a reserve function for the contract owner to hold back some mints from the public sale for some reasons, e.g., personal holdings by team members and future use, etc~\cite{dee_strategies_2022}.
However, suppose there is no requirement to check whether the current supply is beyond the promised limit in the reserve function. When the contract owner seeks to do harm or the private key of the contract owner is leaked, the attacker can achieve unlimited minting to obtain many more NFTs without paying and threaten the NFT rarity.

\vspace{-0.3cm}
\begin{figure}[htb]
\setlength{\abovecaptionskip}{0.cm}
\begin{lstlisting}
function reserveApes() public onlyOwner {
    uint supply = totalSupply();
    uint i;
    for (i = 0; i < 30; i++) {
        _safeMint(msg.sender, supply + i);}}
\end{lstlisting}
\caption{An example of Unlimited Minting defect.}
\label{um}
\end{figure}
\vspace{-0.4cm}

\textbf{Example:} Figure~\ref{um} utilizes a \textit{reserveApes} function for the contract owner to set aside more NFTs through the \textit{\_safeMint} function (line 5) in a smart contract of the famous BAYC NFT project~\cite{smartmud_unlimited_2022}. However, unfortunately, because of the negligence in checking the supply, there can be unlimited minting of the NFTs. The number of minted NFTs (\verb|supply + i|) (line 5) can exceed the max supply of the project by minting 30 more NFTs per batch (lines 4-5). Hence, if the attacker has access to invoke this function and mint unlimited BAYC NFTs for free, the digital scarcity of the NFT can be threatened, and the project will experience huge monetary losses.

\vspace{-0.4cm}
\begin{figure}[htbp]
\setlength{\abovecaptionskip}{0.cm}
\setlength{\belowcaptionskip}{-0.4cm}
\begin{lstlisting}
/* ERC-721 annotations on approve function */
// Throws unless `msg.sender` is the current NFT owner, or an authorized operator of the current owner.
function approve(address to, uint256 tokenId) public virtual override {
    address owner = ERC721.ownerOf(tokenId);
    require(to != owner, "ERC721: approval to current owner");
    /* missing requirement of checking msg.sender */
    _approve(to, tokenId);}
\end{lstlisting}
\caption{An example of Missing Requirements defect.}
\label{mr}
\end{figure}

\textbf{(4) Missing Requirements:}
The ERC-721 standard gives annotations for each function in the interface, helping to ensure the correct operations related to NFTs. Developers should implement those requirements in NFT smart contracts. Unfortunately, the correct logic may fail and harm the NFT owners without requirement checks. For example, if there is no check on the function caller of the \textit{approve} function, anyone can become the operator and thus obtain permission to operate on others' NFTs~\cite{acampana_how_2022}.

\textbf{Example:}
We can find this defect in the \textit{approve} function in Figure~\ref{mr}. The ERC-721 annotates that the approve function should throw an exception unless the function caller is the current NFT owner or an authorized operator (line 2). However, this problematic \textit{approve} function (lines 3-7) does not check the function caller. In this case, if someone calls this function, they can become the operator of the NFT and have permission to transfer it to themselves, thus stealing others' NFTs.

\textbf{(5) Public Burn:}
Burning an NFT is usually carried out to control its price. If the demand is high while the supply is limited, the value may rise as a consequence. Hence, many projects adopt the burning functionality to achieve supply control. Furthermore, NFT burn is also used in other scenarios, such as flaw-fixing and gamification. 
Due to the features of burning, it is important to ensure that only the NFT owner can access the burn function. If anyone can call the burn function to destroy others' NFTs without permission, the project may encounter a tragedy, and all NFTs may evaporate.

\vspace{-0.4cm}
\begin{figure}[htb]
\setlength{\abovecaptionskip}{0.cm}
\begin{lstlisting}
function burn(uint256 tokenId) public {
    _burn(tokenId);}
function _burn(uint256 tokenId) internal virtual {
    address owner = ERC721.ownerOf(tokenId);
    // Clear approvals
    _approve(address(0), tokenId);
    _balances[owner] -= 1;
    delete _owners[tokenId];}
\end{lstlisting}
\caption{An example of Public Burn defect.}
\label{pb}
\end{figure}
\vspace{-0.4cm}

\textbf{Example:}
Figure~\ref{pb} shows that an NFT contract contains a public burn defect. The burn function makes anyone able to burn others' NFTs. The delete statement (line 8) breaks off the ownership between the NFT and its owner. However, there is no check to judge whether the caller of the burn function is the NFT owner, which means anyone can burn others' NFTs.

\para{Code Smell vs. Vulnerability vs. Defect}Code smell is a design flaw that is not wrong but a sign of poor code quality that may be problematic~\cite{beck1999bad}. Vulnerability is a subset of bugs that can be attacked from the perspective of security~\cite{lyu1996handbook}, while a defect is any design flaw in the software that makes it behave incorrectly and not according to its specification~\cite{ieeestandard}.
In other words, the concept \textit{defect} has the widest coverage among these 3 concepts.
In our paper, \textit{Missing Requirements} is a design flaw when developers do not follow the ERC-721 standard. It could cause problems but may not be attacked directly, so it is not a vulnerability. \textit{ERC-721 Reentrancy} contracts can be attacked directly; thus, it is not a code smell. Therefore, we use the concept of defect to unify the 5 issues.
\vspace{-0.5cm}
\section{Methodology}
\label{sec:method}
In this section, we introduce the tool NFTGuard, which can detect the defects defined above. We first give an overview of the approach and then illustrate the details of our detection method from the instruction level and the operational feature level.
\vspace{-0.2cm}
\subsection{Overview}

\begin{figure}[htbp]
\setlength{\abovecaptionskip}{0.1cm}
\setlength{\belowcaptionskip}{-0.7cm}
    \centering
    \includegraphics[width=\linewidth]{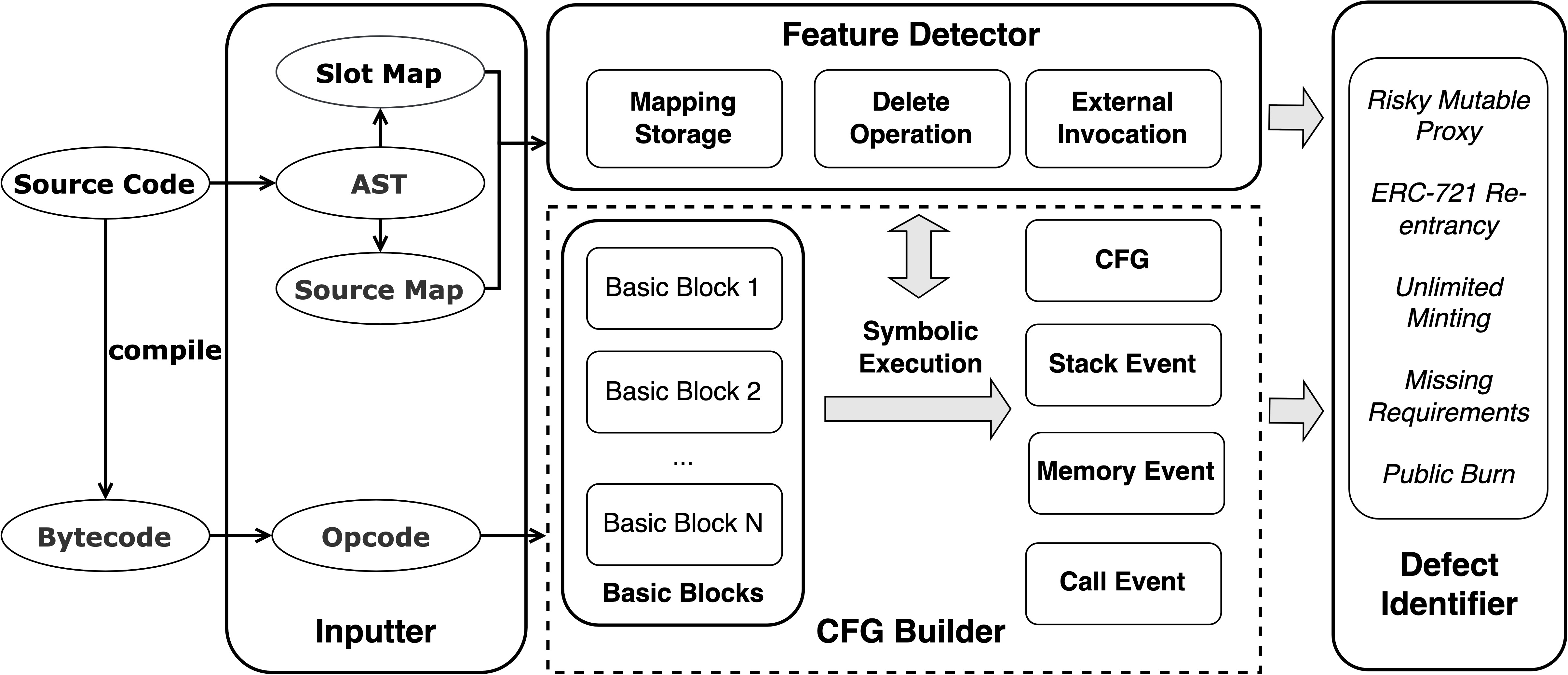}
    \caption{An overview of the approach of NFTGuard.}
    \Description{Overview of the tool.}
    \label{fig:overview}
\end{figure}

Figure~\ref{fig:overview} shows an overview of the NFTGuard approach. There are four main components in our design of NFTGuard, i.e., \textit{Inputter}, \textit{Feature Detector}, \textit{CFG Builder}, and \textit{Defect Identifier}.

Specifically, users need to feed Solidity source code as the input. The contract code will be compiled by the Solidity compiler to obtain the bytecode and AST for further analysis. Specifically, the \textit{Inputter} extracts the source map~\cite{sourcemap} from AST and utilizes a slot map to store the mapping relationship between variables and their slot IDs. The contract bytecode is disassembled into opcodes by utilizing API provided by Geth~\cite{ethereumgo}. Then, NFTGuard builds the CFG (control flow graph) dynamically based on symbolic execution, and critical events (i.e., stack event, memory event, and call event) are recorded for detecting defect features. During the symbolic execution, \textit{Feature Detector} detects three features (i.e., mapping storage, deleted operation, and external invocation). For each defect, NFTGuard designs a specific rule based on the three features and recorded events. Finally, NFTGuard reports the identified defects in the \textit{Defect Identifier} based on the rules.

NFTGuard combines the critical information extracted from the contract source code and the bytecode to help detect defects. 
The motivation for using the source code information is to obtain the slot IDs of state variables and source code when executing a specific opcode from the source map. This can help us to locate defective codes and find defects more effectively in detecting complicated NFT smart contracts with higher code coverage. 
Specifically, in order to find and label those important state variables according to the features of a particular defect, NFTGuard gets the slot IDs of state variables statically and records their names and data types according to the principles~\cite{layout} by analyzing the AST. This helps NFTGuard to monitor the opcodes that handle these variables during symbolic execution. By collaborating with the source map, which bridges the source code and bytecode, NFTGuard can obtain the source code information when executing opcodes. Therefore, NFTGuard can report contract defects based on predesigned feature detection patterns, focusing on unique operations. We give a detailed illustration of these operations in the following paragraphs.

\vspace{-0.2cm}
\subsection{Operational Semantics Modeling}
In this subsection, we model the syntax of several essential instructions, variables, and functions that form the basis of the core analysis module in our detection method.

We first give the operational semantics of some storage- and memory-related instructions as follows:

\noindent Instruction :- $SSTORE(y,z)$ \  write value $y$ to storage address $z$;

\qquad \quad \quad | \ $SLOAD(y,z)$ \  load value $y$ from storage address $z$;

\qquad \quad \quad | \ $MSTORE(y,z)$ \  store value $y$ to memory address $z$;

\qquad \quad \quad | \ $MLOAD(y,z)$ \  load value $y$ from memory address $z$;

\qquad \quad \quad | \ $CALL(t,o,l)$ \ call target function $t$ with arguments loaded starting from memory address $o$ with length $l$.

During the symbolic execution, some critical data structures are updated according to the syntax of EVM instructions. Specifically, $S$ denotes an operand stack, $M$ represents a simulated temporary memory space, $GS$ stands for the stored value of state variables, and the accumulated constraints in the satisfiability modulo theories (SMT) solver~\cite{moura2008z3} during the exploration of paths are defined as the variable $cons$.

As mentioned in the previous subsection, our method utilizes the source code level information to help us find defective codes before and during the symbolic execution. Specifically, by analyzing the AST information, we first get the slot IDs of state variables with their data types. Then, we create a slot map in order to mark state variables and their slot IDs for the analysis of a particular defect. Next, we use a keyword-matching strategy $\delta$ to extract critical variables and their slot IDs by name in the slot map to the set $I_k$ for keyword $k$, e.g., $owner$.

In order to make use of the slot ID information, we define functions $e$ and $i$ to infer the operations related to state variables when tracing paths from the bytecode level. Specifically, the SMT solver accumulates constraints that the symbolic execution framework generates and updates $cons$. To obtain the state variables that determine the condition, we define a function $e$ to extract the storage addresses contained in $cons$ to a set. Furthermore, function $i$ is used to infer slot IDs from read- or write-related addresses in this extracted set.
For instance, suppose $foo$ is a state variable with value type $uint$ and slot ID $0$. The jump condition of judgment $if(foo > 10)$ is $If(ISZERO(GT(10, Ia\_0))$ accumulated in $cons$, and $Ia\_0$ represents the storage address of the state variable with a slot ID of $0$. Function $e$ can be used to extract storage address $Ia\_0$ from $cons$, and function $i$ can infer the slot ID $0$ from $Ia\_0$. It is noticeable that the inference of mapping type state variable is more complicated due to its usage of the hash function, e.g., $KECCAK256$. 

The source map also helps in finding the function-level context during execution. We use the word context to represent the instructions generated by the source code in a function. We can obtain the function name and the corresponding source code from the source map when executing an opcode generated by the function. We use $Context(f)$ to represent the function-level context $f$ when executing a specific opcode.

\vspace{-0.5cm}
\subsection{Operational Features Identification}\label{sec:operations}
Based on the modeling of instruction-level semantics, we put forward the operational features of 3 operations, i.e., Mapping Storage, Delete Operation, and External Invocation. Each contains a series of instructions. During the symbolic execution, these 3 operations reflected in opcode sequences can be utilized to detect defects.

\para{Mapping Storage} The storing of mapping variables consists of some critical operations. Taking \verb|_owners[id] = to| as an example, (1) instructions $MSTORE(id,0)$ and $MSTORE(s,32)$ store values $id$ and the slot ID of variable \verb|_owners| (considered as $s$) to memory address $0$ and $32$, respectively. (2) Then the $KECCAK256$ instruction loads data from memory space $0$ to $64$ and pushes the calculated hash $x$ to the top of the stack. Specifically, this hash represents the storage area of variable \verb|_owners| calculated by $keccak(h(k)\cdot p)$, where $p$ denotes the slot ID, $k$ represents the key of the mapping---$id$, $\cdot$ is concatenation, and $h$ is a function that is applied to the key depending on its type~\cite{layout}. (3) Finally this calculated hash value $x$ is used as the storage address for storing value $to$ by $SSTORE$.

In our method, this feature helps us locate the storing operations of mapping variables and to validate the execution context of functions. Specifically, suppose when executing the opcode $JUMP$, for which the corresponding source code is function \textit{\_mint}, we can know that the symbolic execution is going to execute opcodes in the context of function \textit{\_mint}. Since a standard mint function implements the functionality that stores an address (owner of the token) to a mapping variable's key (token id), as shown in the above example, we can apply rules for identifying the mapping storage when tracing the instructions in the function context. In particular, as we know the $I_{owner}$ before symbolic execution, we can obtain the operations on the state variable \verb|_owners| by identifying its slot ID. With the source map information and operations on the focused state variable, we can infer that the implemented function is a true \textit{\_mint} and the $Context(mint)$ is validated.

\para{Delete Operation} The delete operation is an important symbol of burning tokens in our detection method. Its feature reflected in the opcode sequence is similar to the storage of mapping state variables. However, the difference is that the stored value is the default value of the variable type. Taking \verb|delete bar[1]| for example, \verb|bar| is a $mapping(uint => uint)$ type variable. First, the storage area of \verb|bar[1]| is calculated through $KECCAK256$ by the hash function $keccak((h(1)\cdot p)$. Then a default value of type $uint$---0 is stored at the storage address by $SSTORE$ in the delete operation.

\para{External Invocation} The call operation is relatively complex in Solidity. When executing $CALL$, the stack pops 7 elements. It claims the gas consumed, target address, ether amount to transfer, arguments to pass (from memory), and memory space to store the return data. The prototype reentrancy (e.g., the DAO reentrancy attack~\cite{DAO}), which has attracted much attention, is related to the transferred amount of ether. However, the invocation of an external function is different. The invoked function selector is stored in the memory space loaded by $CALL$ from the stack. Specifically, the memory value read according to the fourth element of the popped elements tells the EVM which function to call during the execution time. The feature of external invocation on a function helps NFTGuard find the existence of \textit{onERC721Received}, which is critical in detecting the \textit{ERC-721 Reentrancy} defect.

\vspace{-0.25cm}
\subsection{Defects Detection}
To detect 5 defects in NFT smart contracts, NFTGuard utilizes a symbolic execution framework to explore contract paths. Predefined models help to recognize the execution states to capture critical features and find defects. In the following paragraphs, we elaborate on the details of finding defects in NFT smart contracts.

\para{(1) Risky Mutable Proxy}
The \textit{Risky Mutable Proxy} defect can be recognized by a proxy method that modifies the state variable storing the proxy's address. So, if there is a proxy address setting function to expose to users, the contract has this kind of defect. We define this rule as below. In this expression, $x$ denotes the user-generated external parameter, and $I_{proxy}$ represents the slot IDs of variables matched by $\delta$ with keyword \textit{proxy}. When $x$ is stored to address $t$, and the inferred slot ID from address $t$ by $i(e)$ is in the set $I_{proxy}$, NFTGuard will report the defect.
\vspace{-0.1cm}
$$RMP \Leftarrow  \dfrac{SSTORE(x,t), i(t)\in I_{proxy}}{x := Input}$$
\vspace{-0.4cm}

\para{(2) ERC-721 Reentrancy}
The memory-related opcodes are focused on finding reentrancy in ``safe'' methods, such as \textit{\_safeMint} and \textit{safeTransferFrom} that contain external function invocations. Specifically, we use the below expression to demonstrate the core idea of identifying a feasible reentrance. 
When executing instruction $CALL$, the argument loaded from memory by the fourth element in the stack $y$ (by $MLOAD$ from memory address $s$) is compared with the left-shifted value $h^{'}$  (by instruction $SHL$) of the function selector of \textit{onERC721Received}---$h$, while $h^{'}$ is stored in the same memory address $s$ by $MSTORE$.
The call target obtained by $y$ should be equal to $h^{'}$ to invoke \textit{onERC721Received}.
Then, the slot IDs extracted from the path conditions are recognized to identify storage addresses by $e$. 
We next feed constraints to the SMT solver and check the satisfaction of the assertion that the state variables are not modified to change the status impeding the reentrance. 
The \textit{ERC-721 Reentrancy} happens when every variable denoted by the storage addresses in $T$ always equals the original value $GS(t)$. 

\vspace{-0.3cm}
$$ER \Leftarrow  \dfrac{CALL(h, y, *), MLOAD(y,s), y \equiv h^{'}, \forall t\in T, t \equiv  GS(t)}{T := e(cons), MSTORE(h^{'},s), S := \langle *,*,*,y,*,*,* \rangle}$$
\vspace{-0.2cm}

\para{(3) Unlimited Minting}
The implemented function \textit{totalSupply} returns the lengths of minted valid tokens and can be used to check the feasibility of minting. If there are no constraints to check the total supply of NFTs and the current minting token number, this contract is considered to contain this kind of defect.

In the expression shown below, $Context(mint)$ stands for the function context obtained from the source map, and the path conditions stored in the symbolic execution states are utilized to find related state variables by $e$. After validating the operation of the mint function by finding mapping storage on variables that store the ownership of an NFT, i.e., $SSTORE(*,t)$, the nonexistence of slot IDs in $I_{supply}$ obtained from $\delta$ and function $i$ represents the \textit{Unlimited Minting}, which means there is no supply check.

\vspace{-0.2cm}
$$UM \Leftarrow  \dfrac{SSTORE(*,t), \forall t\in T, i(t) \notin I_{supply} }{Context(mint), T := e(cons)}$$

\para{(4) Missing Requirements}
The core idea of finding the \textit{Missing Requirements} defect is (1) locating the critical position where we perform the check, and (2) using the SMT solver to check the satisfaction of the requirements at that position. Specifically, for the $approve$ function, the critical position is the operational feature that stores the approved address to the mapping state variable with key---the approved token id. Then, we feed the negated conditions of the correct requirements to the SMT solver and check the feasibility of performing mapping storage in an unintended way. 

The following expression shows our detection logic on Missing Requirements. For an ERC-721 function $f$, the mandated requirements are denoted by a set $E(f)$, where $E_{f_m}$ represents one of the requirements $m$. If any of the negated conditions in $E(f)$, e.g., $\lnot E_{f_m}$, are feasible according to the SMT solver, the contract is considered to have the \textit{Missing Requirements} defect.
\vspace{-0.1cm}
$$MR \Leftarrow  \dfrac{\exists path, \lnot E_{f_m}, E_{f_m} \in E(f) }{Context(f), f \in E}$$
\vspace{-0.3cm}

\para{(5) Public Burn}
To detect the \textit{Public Burn} defect, we should first locate the delete operation in burning NFTs. According to the operational features we define, when the delete operation is recognized, NFTGuard performs a data dependency check. If the caller's address is not one of the decisive variables, the NFT contract has a \textit{Public Burn} defect that misses the caller's address check.
$$PB \Leftarrow  \dfrac{SSTORE(\varnothing,t), \forall t\in T, t \nrightarrow  CALLER }{Context(burn), SLOAD(*,t), T := e(cons)}$$

The above expression represents the features of this kind of defect. First, we identify the function context as $burn$ from the source map; then, we perform detection of the delete feature detection as introduced in the previous subsection. Specifically, we use $\varnothing$ to denote the default value to erase the origin value on address $t$, which is obtained by $SLOAD(*,t)$. It is a \textit{Public Burn} if there are no determining variables such as $msg.sender$ obtained by instruction $CALLER$ in path conditions.
\vspace{-0.15cm}
\section{Experiment}
\label{sec:evaluation}
In this section, we perform a large-scale experiment and give the precision analysis result to measure the effectiveness of NFTGuard based on an open-source dataset.

\vspace{-0.2cm}
\subsection{Experimental Setup}
\label{sec:dataset}
The large-scale experiment was conducted on a server running Ubuntu 18.04.5 LTS and equipped with 12 Intel Xeon Gold 6132 CPUs and 120 GB memory.

\para{Dataset} To identify whether the defined defects are prevalent in real-world Ethereum smart contracts, we utilized an open-source dataset from a GitHub repository~\cite{ContractStatistic} that stores the source code of all verified smart contracts on Etherscan. We downloaded the dataset on August 19th, 2022, and filtered them by the keyword \textit{``NFT''} or \textit{``ERC721''} to extract the NFT smart contracts. As Solidity 0.8.16 was the latest compiler version when writing this paper, we chose this compiler version and removed those contracts that failed to be compiled. Finally, we obtained 16,527 smart contracts and performed the large-scale experiment. It is noticeable that our dataset can better reflect the features of current Solidity smart contracts that are much more complicated than those with lower versions. We compared our dataset with the widely used dataset SmartBugs~\cite{ferreira2020smartbugs}. 
Some key features of these two datasets are shown in Table~\ref{tab:data_features}. It can be seen that smart contracts in our dataset are more complex than those in SmartBugs. Specifically, 100\% of the contracts in our dataset require a Solidity compiler version higher than v0.8.0, while 99.4\% of contracts in SmartBugs use versions lower than v0.5.0. The average lines of code (LOC) and the number of instructions of contracts in our dataset are \textbf{14.1$\times$} and \textbf{5.8$\times$} as much as those in SmartBugs, respectively. The average numbers of public/external functions and state variables claimed in the contracts of our dataset are both nearly \textbf{3$\times$} as much as those in SmartBugs.

\vspace{-0.4cm}
\begin{table}[htbp]
\setlength{\abovecaptionskip}{0.cm}
    \begin{center}
        \caption{Mean of Features of Our Dataset vs. SmartBugs.}
        \label{tab:data_features}
        \resizebox{\columnwidth}{!}{
            \begin{tabular}{c||c|c|c|c}
                \hline
                \diagbox {\textbf{Dataset}}{\textbf{Features}} & \textbf{LOC} & \textbf{\# of Instrs} & \textbf{\# of Funs}  & \textbf{\# of State Vars}  \\
                \hline
                Ours                                           &      1413.6       & 8981.7 & 34.0                                & 19.7                                      \\
                \hline
                SmartBugs                                      &       99.9       &  1545.5   &     12.5                         &            6.6                                    \\
                \hline
            \end{tabular}
        }
    \end{center}
\end{table}
\vspace{-0.4cm}

\para{Evaluation Metrics} We summarize the following research questions (RQ) to evaluate the effectiveness of NFTGuard.

\begin{enumerate}[RQ1.]
	\item How effective is NFTGuard in detecting the 5 defined defects in our dataset? Can NFTGuard really find those defects?
	\item In terms of effectiveness, how is the precision of finding defects in NFT smart contracts?
        \item Can NFTGuard find NFT-related defects that other tools are unable to detect?
\end{enumerate}

\vspace{-0.25cm}
\subsection{Answer to RQ1: Defects Detection in a Large-Scale Dataset}
To answer RQ1, we ran NFTGuard on the source code of the 16,527 verified smart contracts in our dataset. The experimental results given in Table~\ref{tab:nft_defect} (the second and the third columns) show the frequency of each defect in NFT contracts on Ethereum. Our NFTGuard only identifies whether the contract contains a defect, so if the same kind of defect occurs multiple times, we only count it once.

\vspace{-0.4cm}
\begin{table}[htbp]
\setlength{\abovecaptionskip}{0cm}
    \begin{center}
        \caption{Defects in NFT Contracts and Precision of NFTGuard.}
        \label{tab:nft_defect}
        \resizebox{\columnwidth}{!}{
        \begin{tabular}{l||c|c|c|c|c|c}
            \hline
            \textbf{Contract Defect}    & \textbf{\# Defects} & \textbf{Per(\%)} & \# \textbf{Samples} & \textbf{\# TP} & \textbf{\# FP} & \textbf{Prec(\%)} \\
            \hline
            \textit{Risky Mutable Proxy}                  & 15             & 0.09  &  15 & 15 & 0 & 100.0         \\
            \hline
            \textit{ERC-721 Reentrancy}           & 503             & 3.0   & 81 & 66 & 15 & 81.5          \\
            \hline
            \textit{Unlimited Minting}            & 781              & 4.7    & 86 & 84 & 2 & 97.7           \\
            \hline
            \textit{Missing Requirements}           & 81               & 0.49  & 44 & 44 & 0 & 100.0           \\
            \hline
            \textit{Public Burn}           & 44               & 0.27  & 30 & 28 & 2 & 93.3           \\
            \hline
        \end{tabular}
        }
    \end{center}
\end{table}
\vspace{-0.4cm}

\textit{Unlimited Minting} is the most frequent kind of defect in our dataset, and approximately 4.7\% of the NFT smart contracts contain this defect. A total of 3.0\% of NFT smart contracts contain \textit{ERC-721 Reentrancy} defects. Furthermore, the percentage of \textit{Risky Mutable Proxy}, \textit{Missing Requirements}, and \textit{Public Burn} defects are all lower than 1\%, with 15 (0.09\%), 81 (0.49\%), and 44 (0.27\%) NFT smart contracts containing this kind of defect, respectively.

In addition, the experimental results showed that 4 NFT smart contracts contain 3 kinds of the 5 defined defects, and 85 NFT smart contracts contain 2 kinds of defined defects. Furthermore, there are 1,331 NFT smart contracts in our dataset containing at least one kind of defect, meaning approximately 8.1\% of NFT smart contracts have some defects, as reported by NFTGuard.

\vspace{-0.2cm}
\subsection{Answer to RQ2: Evaluation of NFTGuard}
To answer RQ2, we evaluate the performance of NFTGuard in this subsection. 
Specifically, we randomly sample a number of contracts from reported positives in our detection results for each defect. To establish the sample size per defect, we follow the sampling method based on a confidence interval~\cite{confidenceinterval} for generalization to the population of the total number of issues found for that defect.
We set a confidence interval of 10 and a confidence level of 95\% and calculate the number of samples we need to collect~\cite{confidenceintervalcalculator}. 
The calculated results of the five defects are 13, 81, 86, 44, and 30, respectively.
Then, the evaluation dataset is sampled according to the results and manually labeled carefully by two of the authors.
In the labeling process, we separate the false positives and true positives to analyze the performance of NFTGuard. This approach has also been adopted by other related works~\cite{kalra2018zeus,luu2016making,jiang2018contractfuzzer}.

The fourth to seventh columns of Table~\ref{tab:nft_defect} summarize the results of applying NFTGuard to our labeled samples (we analyzed all the 15 contracts with \textit{Risky Mutable Proxy} defects to make the results more reliable). The fifth and sixth columns show the number of samples separated into true positives (TP) and false positives (FP). We use the precision rate to show the performance when detecting each defect. The precision rate can be calculated as $\frac{\#TP}{\#TP+\#FP}\times 100\%$. Furthermore, we also calculate the overall precision to show the effectiveness of NFTGuard.
The overall result is calculated by $\frac{\sum_{i = 1}^{n}p_{c_i}\times|c_i|}{\sum_{i = 1}^{n}|c_i|}$, in which $p_{c_i}$ represents the precision of detecting defect $i$, and $|c_i|$ is the number of defect $i$ in our dataset.

The detection of smart contracts containing \textit{Risky Mutable Proxy} and \textit{Missing Requirements} achieves a precision rate of 100\%. For \textit{ERC-721 Reentrancy}, \textit{Unlimited Minting}, and \textit{Public Burn},  NFTGuard reports them at a precision of 81.5\%, 97.7\%, and 93.3\%, respectively. Furthermore, the overall precision of NFTGuard achieves 92.6\%. 

\para{False positives} There are some false positives in our experimental results. (1) For \textit{ERC-721 Reentrancy}, we found that there is no change in the state variable that determines the path condition to execute the instruction $CALL$ that invokes the external function. For example, in the function containing a safe function, there is only one statement \textit{\_safeMint} or \textit{safeTransferFrom} with some \textit{require} statements before it. The intent of the function is not to modify a state variable that is used to prevent the caller from calling it again. Hence, it is feasible to reenter the function, but the effect achieved by reentrancy is the same as calling it again in another transaction.
(2) For \textit{Public Burn}, NFTGuard needs to compare the \textit{msg.sender} to the NFT owner's address, e.g., \verb|msg.sender == ownerOf(id)|. However, a mapping variable might return the compared result directly. For example, the mapping variable \verb|addr2owner[msg.sender])| returns a boolean value $true$ or $false$ to represent whether the \textit{msg.sender} is the owner of the NFT. We cannot know the details of the mapping variable, thus leading to false positives.
(3) As for \textit{Unlimited Minting}, false positives are those contracts using constant values which are hardcoded in the EVM bytecode to check the mint quantity. As variables claimed as constants do not have slot IDs, NFTGuard fails to detect the comparison operation and leads to errors.

\para{False Negatives}To find contracts with defects that NFTGuard failed to report, we follow the same sampling method that is utilized to analyze the precision. We randomly sampled 95 contracts from 15,196 contracts where no defect was reported with a confidence interval of 10 and a confidence level of 95\%. We then manually label them to find false negatives that NFTGuard missed. In total, we find 5 of 95 contracts are false negatives, and all of them indeed contain \textit{ERC-721 Reentrancy} defects without being reported.
All of the 5 false negatives are caused by the path explosion during the symbolic execution. Specifically, these contracts contain a large number of branches in their CFG, which leads to huge search space.
In the design of NFTGuard, to avoid path explosions, we limited the maximum loop times, the depth of path searching, and the time of execution; thus, the reentrancy point is not checked by NFTGuard. For instance, there are 6 \textit{require} statements (lines 1313-1322) in total before the nested \textit{for loop} (lines 1324-1326) of minting NFTs in contract\footnote{\href{https://etherscan.io/address/0x1364748db21598e16c76d5992c82524b5f8b4099\#code}{https://etherscan.io/address/0x1364748db21598e16c76d5992c82524b5f8b4099\#code}}, one of the 5 false negatives, which makes the searching paths so deep that failed to be detected by NFTGuard.

\para{Contracts containing \textit{Unlimited Minting} defect} From the large-scale experimental results, the number of contracts reported to contain \textit{Unlimited Minting} defect is far greater than the number of other kinds of defects. We found that many projects provide a reserve function to set aside NFTs without checking the max supply of the project, which is a design flaw when implementing NFT smart contracts. From what we are concerned, the project may not be attacked directly by the project team itself because they can not make more profits by reserving many more NFTs than the normal sale. However, we cannot overlook the possibility of the leakage of the contract owner's private key, which the attacker can exploit to shut down the project and cause huge economic losses.

\para{New defective code designs} In our labeling process, we found that NFTGuard detected some defective code designs which have not been reported. Specifically, some contracts use a bool type variable to judge whether the caller of function \textit{burn} should be checked, but the user can set this variable. Furthermore, some contracts dismiss the required zero address check when implementing function \textit{\_mint} even though the annotation mandates it.

\vspace{-0.15cm}
\subsection{Answer to RQ3: Comparison Experiment\label{sec:comparison}}
In RQ3, we compare NFTGuard with other tools and illustrate why NFTGuard can find more NFT defects than them.

We first collect a list of vulnerability detection tools for smart contracts from papers of top journals/conferences on software and security, e.g., ISSTA, ICSE, S\&P, etc., and Mythril~\cite{mythril}, which is officially recommended by Ethereum.
We find that none of them support detecting the 4 NFT-related defects we newly defined in this paper, except for \textit{Reentrancy}. 
To evaluate whether they can be used to detect \textit{ERC-721 Reentrancy}, we then choose 6 baseline tools for our comparison experiment: Mythril, Oyente~\cite{luu2016making}, Sailfish~\cite{rao2012sailfish}, Securify1~\cite{tsankov2018securify}, Securify2~\cite{securify2}, and Smartian~\cite{choi2021smartian} considering from 4 aspects: (1) availability of tool source code; (2) usability of command-line interface for large-scale experiment; (3) supporting Solidity source code; (4) ability to report defective code location for manual examination. We use these 6 tools to analyze the randomly select
1,000 smart contracts from our dataset of 16,527 contracts.

The results show that NFTGuard reports 14 contracts with \textit{ERC-721 Reentrancy} defects, while none are reported by any of the 6 tools. Furthermore, the results show these tools cannot analyze many NFT contracts and generate compilation errors or timeouts with no result.
In addition, we find that although these tools can detect \textit{Reentrancy} issue, they only focus on detecting code patterns that are used to transfer Ethers, e.g., \textit{call.value()}. However, the \textit{ERC-721 Reentrancy} is only related to token transfer with external function invocation and does not involve Ether transfer; thus, previous tools cannot detect \textit{ERC-721 Reentrancy}.

\vspace{-0.1cm}
\section{Discussion}
\label{sec:discussion}
\subsection{Case Study}
We give an example to show how the attacker exploited the defect and the importance of detecting these defects in NFT smart contracts reported by NFTGuard.

\begin{figure}[htbp]
\setlength{\abovecaptionskip}{0cm}
\setlength{\belowcaptionskip}{-0.8cm}
    \centering
    \includegraphics[width=3in]{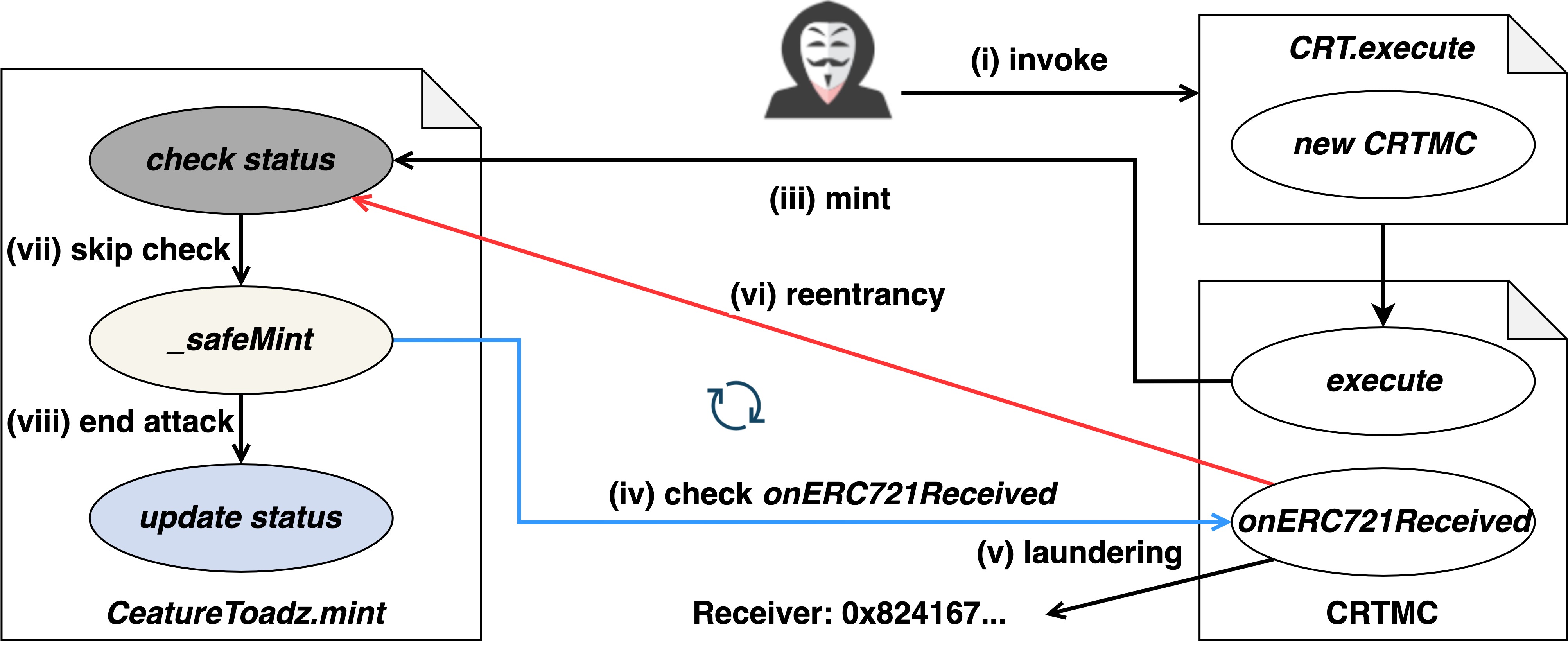}
    \caption{Illustration of the ERC-721 Reentrancy case.}
    \Description{Illustration of the ERC-721 Reentrancy case.}
    \label{fig:case_illustrate}
\end{figure}

There is an \textit{ERC-721 Reentrancy} defect in the contract\footnote{\href{https://etherscan.io/token/0xa4631a191044096834ce65d1ee86b16b171d8080\#code}{https://etherscan.io/token/0xa4631a191044096834ce65d1ee86b16b171d8080\#code}}. The project contains 8,888 NFTs, which was worth approximately 2.65 million USD by Nov. 2022.
Figure~\ref{fig:case_illustrate} illustrate the attack process of the attacking transaction\footnote{\href{https://etherscan.io/tx/0x4f09884136010cee2c56ff646ba2a5d3a030a5f180e505200f991eef8aa91207}{https://etherscan.io/tx/0x4f09884136010cee2c56ff646ba2a5d3a030a5f180e505200f991\ eef8aa91207}} that was launched by an attacking contract\footnote{\href{https://etherscan.io/address/0xdFF832F6988E4a9E3FCfBfF4cc24d052143aba0E\#code}{https://etherscan.io/address/0xdFF832F6988E4a9E3FCfBfF4cc24d052143aba0E\#code}}. Figure~\ref{case_code} shows a simplified code snippet of the defective contract for illustration.
The attacking contract \textit{CRT} first created a contract \textit{CRTMC} to invoke function \textit{CreatureToadz.mint} through function \textit{execute} to buy NFTs (step \textit{iii}). In the \textit{onERC721Received} function, the created contract reenters into the defective contract and skips the status check (line 2) to mint more NFTs (step \textit{vi}) before the minted status was modified (line 6). Notably, the minted NFT in this process is transferred (step \textit{v}) to another address. It causes the \verb|senderBalance| of \textit{CRTMC} to remain at zero and skips the balance check (line 4). 
In this transaction, the attacker minted 25 NFTs, which is much larger than threshold 1. The attacker violated the intent of the function and undermined the software fairness, which impacted the NFT rarity, hurting other buyers.

\vspace{-0.2cm}

\begin{figure}[htb]
\setlength{\abovecaptionskip}{0cm}
\setlength{\belowcaptionskip}{-0.7cm}
\begin{lstlisting}
function mint(address _wallet, uint256 _mintAmount) public payable {
    require(mintlist[_wallet] != true, "You've already minted CreatureToadz! Don't be greedy!");
    // maxMintAmount is set to constant 1
    if (senderBalance > 0) {require(_mintAmount + senderBalance <= maxMintAmount, "You can only mint one CreatureToadz!");}
    // safeMint that invokes onERC721Received
    mintlist[_wallet] = true;}
\end{lstlisting}
\caption{Code snippet of the defective CreatureToadz.}
\label{case_code}
\end{figure}


\subsection{Limitation} 
\vspace{-0.1cm}
\para{Limitations of the approach/tool} From the perspective of our approach/tool, we find that, currently, NFT smart contracts are much more complicated in terms of the various features we show in Section~\ref{sec:dataset}. This makes the symbolic execution process time-consuming. In the experiment on our dataset, NFTGuard took 469.8 seconds per NFT smart contract on average, due to the limitation brought by a large number of opcodes and deep function-level invocations that lead to path explosion.
Furthermore, NFTGuard currently supports the detection of 5 types of defects. New emerging defects may not be covered due to the limitation of our particular design on each defect. However, NFTGuard can be extended to detect more types of new defects in future work.

\para{Threats to the validity of the experiment} From the perspective of our experiment, the manual labeling process may also lead to mistakes in dividing the false negatives and true negatives. However, we use a double-check process to address the problem and update the labeled dataset in a timely manner to ensure correctness.
Furthermore, the NFT contracts in our dataset are filtered by the keywords \textit{``NFT''} or \textit{``ERC721''} from the source code, which may include some contracts that are not related to NFT. However, the manually labeled contracts in our dataset are all NFT smart contracts. Therefore, we believe that this does not influence our experiment.
The other threat is that the NFT smart contracts deployed on Ethereum are growing quickly, and a huge number of these NFT smart contracts are not detected in our dataset. With the news reporting unsafe coding samples to developers, the number of defects in NFT smart contracts may decrease, leading to different results from the findings in our paper.

Despite the previous limitations, NFTGuard finds many defective NFT smart contracts that have not been reported but can do great harm to the security of the NFT ecosystem with high overall precision. Since most of the NFT projects publicize their NFT smart contracts' source code to make themselves trustworthy,  both contract developers and third-party authorities can utilize NFTGuard to inspect the security of NFT contracts before and after deployment.

\vspace{-0.2cm}
\subsection{Extension for ERC-1155}
In the paper, we only focus on ERC-721 NFTs. The ERC-1155 Multi Token Standard~\cite{eip1155} is another official standard for NFT provided on EIPs, which makes contracts capable of managing multiple token types, such as ERC-20 and ERC721. 
The ERC-1155 makes it possible to transfer multiple token types at once, including ERC721 tokens, thereby enabling token transfers with lower transaction costs.
Specifically, the ERC-721 standard uses a one-dimensional mapping to represent a single non-fungible index, and each collection of NFTs requires deploying a separate contract. In comparison, the ERC-1155 standard uses a two-dimensional mapping that allows for each token ID to represent a new configurable token type, e.g., NFT. For example, two-dimensional mappings \verb|balances[a][c]| and \verb|balances[b][c]| can represent the account balance of user $c$ in two different NFT collections $a$ and $b$.

Therefore, the ERC-1155 provides the extension of the ERC-721 that combines different NFT collections into a single contract, and each collection represented by a token ID still keeps the features enabled by ERC-721. Specifically, 
as multi-dimensional arrays and mappings are adopted to represent the multi-token, detection patterns based on one-dimensional mapping type variables, e.g., \verb|balances[to]|, should be extended to multi-dimensional mappings, e.g., \verb|balances[id][account]|.
Specifically, two-time hash calculation by \textit{keccak} should be identified. The keccak hash of 
\verb|balance[id]| is used again as the parameter of \textit{keccak} to find the location of \verb|balances[id][account]|, recursively. Therefore, the features of mapping storage and delete operation can be extended by identifying the calculation of two-iteration hash from the events of stack and memory during execution. In addition, the external invocation can be extended by identifying call events on the function selector hash of \textit{onERC1155Received} instead of \textit{onERC721Received}.

\vspace{-0.2cm}
\subsection{Possible Solutions for NFT Defects}
To ensure NFT contracts' security, we not only design NFTGuard to detect potential defects but also want to help developers write secure NFT smart contracts. Therefore, we provide possible suggestions for developers to avoid involving the defined defects in NFT contracts in this subsection.

Table~\ref{tab:solutions} shows a brief possible solution for each defect. Due to the page limitation, we give two examples of applying the suggested solutions for \textit{Public Burn} (lines 1-4) and \textit{Unlimited Minting} (lines 5-11), respectively, in Figure~\ref{solution}. More examples can be found in our repository. Specifically, for \textit{Public Burn}, we add a permission check (line 3) to impede the unauthorized burning of tokens by the delete operation. For \textit{Unlimited Minting}, we add a comparison of the mint amount with the max supply before minting (line 8).
Furthermore. for \textit{Risky Mutable Proxy}, we proposed 2 solutions, i.e., (1) set proxy address with a constant value without any setter function for this variable, and (2) use the parameter passed by the constructor to set this address. Therefore, neither the public nor the contract owner could change the proxy account address.
For \textit{ERC-721 Reentrancy}, it is suggested that (1) put change operations of the state variable that determines the condition to mint NFTs before methods that invoke \textit{onERC721Received}, e.g., \textit{\_safeMint}, or (2) use reentrant locks such as ReentrancyGuard~\cite{reenguard}. 
For \textit{Missing Requirements}, developers should follow the annotations of every interface and check the implementation of the mandated requirements.

\vspace{-0.4cm}
\begin{table}[htbp]
    \setlength\tabcolsep{3pt}
    \setlength{\abovecaptionskip}{0.cm}
    \begin{center}
        \caption{Possible Solutions for the 5 Defects.}
        \label{tab:solutions}
        \resizebox{\columnwidth}{!}{
        \begin{tabular}{l||p{5cm}}
            \hline
            \textbf{Contract Defect} & \textbf{Possible Solution}  \\
            \hline
            \textit{Risky Mutable Proxy}               &  Hardcode address or set it in the constructor without any setter function.                  \\
            \hline
            \textit{ERC-721 Reentrancy}        & Use reentrant locks or advance the modification of deterministic variables.              \\
            \hline
            \textit{Unlimited Minting}         & Check mint amount with the max supply before minting.         \\
            \hline
            \textit{Missing Requirements}        &  Review code implementation according to mandated requirements.            \\
            \hline
            \textit{Public Burn}        & Check the caller's permission before burning NFTs.                    \\
             \hline
        \end{tabular}
        }
    \end{center}
\end{table}
\vspace{-0.6cm}
\vspace{-0.3cm}
\begin{figure}[htbp]
\setlength{\abovecaptionskip}{0.cm}
\begin{lstlisting}
function burn(uint256 tokenId) public {
    // fix public burn by permission checking
    require(msg.sender == ownerOf(tokenId));
    _burn(tokenId);}
function reserveApes() public onlyOwner {
    uint supply = totalSupply();
    // fix unlimited minting by comparing with max supply
    require(supply + 30 < maxSupply)
    uint i;
    for (i = 0; i < 30; i++) {
        _safeMint(msg.sender, supply + i);}}
\end{lstlisting}
\caption{Examples of applying the suggested solutions.}
\label{solution}
\end{figure}
\vspace{-0.4cm}

To show the effectiveness of our solutions, we randomly choose 10 defective contracts for each defect that are correctly detected by NFTGuard and apply the suggestions to fix them. Then, we use NFTGuard to analyze the fixed contracts, and the results show that NFTGuard does not report any defects.

\vspace{-0.15cm}
\section{Related Work}
\vspace{-0.1cm}
\label{sec:rw}
\para{Smart contracts defects}
Chen et al. proposed the first work to define smart contract defects on Ethereum, considering the practitioner's perspective~\cite{chen2020defining}. They collected posts from StackExchange and used open card sorting to find and classify 20 contract defects. They designed a survey to collect developers' feedback, voicing their concerns about these defects. Furthermore, to detect the defects they defined, they also developed a tool named DefectChecker in another work~\cite{chen2021defectchecker} by analyzing the contracts' bytecode. However, their defined contract defects cannot cover new scenarios of security problems in NFT smart contracts. Specifically, although defect \textit{Reentrancy} is also defined in their work, it is not the same as ours according to unique features brought by the NFT as illustrated in Section~\ref{sec:method} and experimental results shown in Section~\ref{sec:comparison}.

\para{Detection tools for security problems on smart contracts}
Recently, many program analysis tools have been developed to focus on detecting security problems on Solidity smart contracts. Specifically, 
Luu et al. proposed the first symbolic execution tool Oyente~\cite{luu2016making}. They utilized an SMT solver Z3 to execute instructions and try to obtain the complete control flow graph and possible paths. Furthermore, MAIAN~\cite{nikolic2018finding}, Securify~\cite{tsankov2018securify}, Osiris~\cite{torres2018osiris}, Ethainter~\cite{brent2020ethainter}, Sailfish~\cite{rao2012sailfish}, Mythril~\cite{mythril}, and Slither~\cite{feist2019slither} are also static analysis tools for detecting vulnerabilities in Solidity smart contracts. Other works such as ContractFuzzer~\cite{jiang2018contractfuzzer}, sFuzz~\cite{nguyen2020sfuzz}, Smartian~\cite{choi2021smartian}, and Echidna~\cite{grieco2020echidna} are based on dynamic testing and analysis. However, all of these tools do not perform well when detecting NFT-related security problems in smart contracts.

\para{Security analysis on NFT ecosystem}
With the popularity of the NFT ecosystem, there have been many works that aim to analyze its security. Wang et al. proposed the first systematic study on the NFT ecosystem~\cite{wang2021non}, and they gave a security evolution of NFT solutions from the perspective of their design models, opportunities, and challenges. Das et al. study the security, privacy, and usability issues in the NFT ecosystem~\cite{das2021understanding}. They provided an overview of how the NFT ecosystem works and analyzed related data to quantify malicious trading behaviors. Furthermore, Wachter et al. investigated NFT wash trading risks from the transaction level to analyze the security of the NFT ecosystem~\cite{siakam_nft_2022}.

However, although some of these previous studies mention the underlying risks of NFT smart contracts, they do not analyze what they are and how they influence the security of the NFT ecosystem. Furthermore, there are no defined contract defects for NFT smart contracts, which threatens the whole NFT ecosystem.
In our work, we not only define contract defects in NFT smart contracts, discuss their impacts and provide possible solutions, but also develop a tool to expose defective NFT contracts to protect the NFT ecosystem.




\vspace{-0.2cm}
\section{Conclusion}
\label{sec:conclusion}
There are two main parts in this paper---the definition and detection of defects. For the definition of defects, we first collected posts on StackOverflow and security analysis reports related to security issues in NFT smart contracts that had been encountered by developers or were considered highly risky. Then we manually analyzed these posts and reports, and defined 5 defects in NFT smart contracts. Furthermore, we give a code example and possible solutions for each defect. To find the security problems in real-world contracts, we developed NFTGuard, a tool based on symbolic execution to detect the 5 defined defects in NFT smart contracts.

NFTGuard combines the source code level information with bytecode. Specifically, NFTGuard extracts the storage slot of some critical state variables in judging problematic operations that may lead to defects on those state variables before symbolic execution. While executing the instructions, NFTGuard matches key operations, i.e., mapping storage, the delete operation, and external invocation, and reports defects according to the predefined defect patterns in collaboration with the source map. NFTGuard focuses on the loading and storing of state variables and the 3 critical operations to perform the detection. In addition, NFTGuard supports the latest compiler versions, e.g., v0.8+, and is also extensible for developers to program other detection patterns to identify more defects with high code coverage.
The experimental results show that NFTGuard finds 1,331 smart contracts in the dataset to contain at least one of our defined defects. Furthermore, the overall precision achieved by NFTGuard is 92.6\%.


\begin{acks}
This work is partially supported by fundings from the National Key R\&D Program of China (2022YFB2702203), the National Natural Science Foundation of China (No. 62032025), and Technology Program of Guangzhou, China (No. 202103050004).
\end{acks}

\normalem

\bibliography{ref}


\begin{thebibliography}{58}


\ifx \showCODEN    \undefined \def \showCODEN     #1{\unskip}     \fi
\ifx \showDOI      \undefined \def \showDOI       #1{#1}\fi
\ifx \showISBNx    \undefined \def \showISBNx     #1{\unskip}     \fi
\ifx \showISBNxiii \undefined \def \showISBNxiii  #1{\unskip}     \fi
\ifx \showISSN     \undefined \def \showISSN      #1{\unskip}     \fi
\ifx \showLCCN     \undefined \def \showLCCN      #1{\unskip}     \fi
\ifx \shownote     \undefined \def \shownote      #1{#1}          \fi
\ifx \showarticletitle \undefined \def \showarticletitle #1{#1}   \fi
\ifx \showURL      \undefined \def \showURL       {\relax}        \fi
\providecommand\bibfield[2]{#2}
\providecommand\bibinfo[2]{#2}
\providecommand\natexlab[1]{#1}
\providecommand\showeprint[2][]{arXiv:#2}

\bibitem[iee(1994)]%
        {ieeestandard}
 \bibinfo{year}{1994}\natexlab{}.
\newblock \bibinfo{title}{IEEE Standards Collection for Software Engineering}.
\newblock
\newblock


\bibitem[eth(2022)]%
        {ethereumgo}
 \bibinfo{year}{2022}\natexlab{}.
\newblock \bibinfo{title}{ethereum/go-ethereum}.
\newblock
\newblock
\urldef\tempurl%
\url{https://github.com/ethereum/go-ethereum}
\showURL{%
\tempurl}


\bibitem[omn(2022)]%
        {omni_hacker}
 \bibinfo{year}{2022}\natexlab{}.
\newblock \bibinfo{title}{Hacker drains \$1.4 million worth of {ETH} from {NFT}
  lender {Omni}}.
\newblock
\newblock
\urldef\tempurl%
\url{https://www.theblock.co/post/156800/hacker-drains-1-4-million-worth-of-eth-from-nft-lender-omni}
\showURL{%
\tempurl}
\newblock
\shownote{Section: Hacks}.


\bibitem[lay(2022)]%
        {layout}
 \bibinfo{year}{2022}\natexlab{}.
\newblock \bibinfo{title}{Layout of {State} {Variables} in {Storage} —
  {Solidity} 0.8.16 documentation}.
\newblock
\newblock
\urldef\tempurl%
\url{https://docs.soliditylang.org/en/v0.8.16/internals/layout_in_storage.html}
\showURL{%
\tempurl}


\bibitem[med(2022)]%
        {medium}
 \bibinfo{year}{2022}\natexlab{}.
\newblock \bibinfo{title}{Medium – {Where} good ideas find you.}
\newblock
\newblock
\urldef\tempurl%
\url{https://medium.com/}
\showURL{%
\tempurl}


\bibitem[ree(2022)]%
        {reenguard}
 \bibinfo{year}{2022}\natexlab{}.
\newblock \bibinfo{title}{{OpenZeppelin} - {ReentrancyGuard}}.
\newblock
\newblock
\urldef\tempurl%
\url{https://github.com/OpenZeppelin/openzeppelin-contracts/blob/36951d58386b9fee81b237e6c6626c9115ccef3a/contracts/security/ReentrancyGuard.sol}
\showURL{%
\tempurl}


\bibitem[pec(2022)]%
        {peckshield}
 \bibinfo{year}{2022}\natexlab{}.
\newblock \bibinfo{title}{{PeckShield}}.
\newblock
\newblock
\urldef\tempurl%
\url{https://peckshield.com/#home}
\showURL{%
\tempurl}


\bibitem[slo(2022)]%
        {slowmist}
 \bibinfo{year}{2022}\natexlab{}.
\newblock \bibinfo{title}{Slowmist}.
\newblock
\newblock
\urldef\tempurl%
\url{https://www.slowmist.com/}
\showURL{%
\tempurl}


\bibitem[Con(2022)]%
        {ContractStatistic}
 \bibinfo{year}{2022}\natexlab{}.
\newblock \showarticletitle{Smart contract statistic}.
\newblock \bibinfo{journal}{\emph{Web:
  https://github.com/tintinweb/smart-contract-sanctuary. [Accessed:
  19-August-2022]}} (\bibinfo{year}{2022}).
\newblock


\bibitem[sol(2022)]%
        {solidity}
 \bibinfo{year}{2022}\natexlab{}.
\newblock \bibinfo{title}{{Solidity} 0.8.16 documentation}.
\newblock
\newblock
\urldef\tempurl%
\url{https://docs.soliditylang.org/en/v0.8.16/}
\showURL{%
\tempurl}


\bibitem[sou(2022)]%
        {sourcemap}
 \bibinfo{year}{2022}\natexlab{}.
\newblock \bibinfo{title}{Source {Mappings} — {Solidity} 0.8.16
  documentation}.
\newblock
\newblock
\urldef\tempurl%
\url{https://docs.soliditylang.org/en/v0.8.16/internals/source_mappings.html}
\showURL{%
\tempurl}


\bibitem[sta(2022)]%
        {stackoverflow}
 \bibinfo{year}{2022}\natexlab{}.
\newblock \bibinfo{title}{Stack {Overflow} - {Where} {Developers} {Learn},
  {Share}, \& {Build} {Careers}}.
\newblock
\newblock
\urldef\tempurl%
\url{https://stackoverflow.com/}
\showURL{%
\tempurl}


\bibitem[twi(2022)]%
        {twitter}
 \bibinfo{year}{2022}\natexlab{}.
\newblock \bibinfo{title}{Twitter}.
\newblock
\newblock
\urldef\tempurl%
\url{https://twitter.com/}
\showURL{%
\tempurl}


\bibitem[wyv(2022)]%
        {wyvern}
 \bibinfo{year}{2022}\natexlab{}.
\newblock \bibinfo{title}{Wyvern {Protocol}}.
\newblock
\newblock
\urldef\tempurl%
\url{https://wyvernprotocol.com/}
\showURL{%
\tempurl}


\bibitem[acampana(2022)]%
        {acampana_how_2022}
\bibfield{author}{\bibinfo{person}{acampana}.} \bibinfo{year}{2022}\natexlab{}.
\newblock \bibinfo{title}{How to transfer a {NFT} from one account to another
  using {ERC721}?}
\newblock
\newblock
\urldef\tempurl%
\url{https://stackoverflow.com/q/67317392}
\showURL{%
\tempurl}


\bibitem[Beck et~al\mbox{.}(1999)]%
        {beck1999bad}
\bibfield{author}{\bibinfo{person}{Kent Beck}, \bibinfo{person}{Martin Fowler},
  {and} \bibinfo{person}{Grandma Beck}.} \bibinfo{year}{1999}\natexlab{}.
\newblock \showarticletitle{Bad smells in code}.
\newblock \bibinfo{journal}{\emph{Refactoring: Improving the design of existing
  code}} \bibinfo{volume}{1}, \bibinfo{number}{1999} (\bibinfo{year}{1999}),
  \bibinfo{pages}{75--88}.
\newblock


\bibitem[BlockSec(2022a)]%
        {blocksec_how_2022}
\bibfield{author}{\bibinfo{person}{BlockSec}.}
  \bibinfo{year}{2022}\natexlab{a}.
\newblock \bibinfo{title}{How {Akutar} {NFT} loses {34M} {USD}}.
\newblock
\newblock
\urldef\tempurl%
\url{https://blocksecteam.medium.com/how-akutar-nft-loses-34m-usd-60d6cb053dff}
\showURL{%
\tempurl}


\bibitem[BlockSec(2022b)]%
        {blocksec_safemint}
\bibfield{author}{\bibinfo{person}{BlockSec}.}
  \bibinfo{year}{2022}\natexlab{b}.
\newblock \bibinfo{title}{When “{SafeMint}” {Becomes} {Unsafe}: {Lessons}
  from the {HypeBears} {Security} {Incident}}.
\newblock
\newblock
\urldef\tempurl%
\url{https://blocksecteam.medium.com/when-safemint-becomes-unsafe-lessons-from-the-hypebears-security-incident-2965209bda2a}
\showURL{%
\tempurl}


\bibitem[Brekke and Fischer(2021)]%
        {brekke_digital_2021}
\bibfield{author}{\bibinfo{person}{Jaya~Klara Brekke} {and}
  \bibinfo{person}{Aron Fischer}.} \bibinfo{year}{2021}\natexlab{}.
\newblock \showarticletitle{Digital scarcity}.
\newblock \bibinfo{journal}{\emph{Internet Policy Review}}
  \bibinfo{volume}{10}, \bibinfo{number}{2} (\bibinfo{date}{April}
  \bibinfo{year}{2021}).
\newblock
\showISSN{2197-6775}
\urldef\tempurl%
\url{https://policyreview.info/glossary/digital-scarcity}
\showURL{%
\tempurl}


\bibitem[Brent et~al\mbox{.}(2020)]%
        {brent2020ethainter}
\bibfield{author}{\bibinfo{person}{Lexi Brent}, \bibinfo{person}{Neville
  Grech}, \bibinfo{person}{Sifis Lagouvardos}, \bibinfo{person}{Bernhard
  Scholz}, {and} \bibinfo{person}{Yannis Smaragdakis}.}
  \bibinfo{year}{2020}\natexlab{}.
\newblock \showarticletitle{Ethainter: a smart contract security analyzer for
  composite vulnerabilities}. In \bibinfo{booktitle}{\emph{Proceedings of the
  41st ACM SIGPLAN Conference on Programming Language Design and
  Implementation}}. \bibinfo{pages}{454--469}.
\newblock


\bibitem[Chen et~al\mbox{.}(2020)]%
        {chen2020defining}
\bibfield{author}{\bibinfo{person}{Jiachi Chen}, \bibinfo{person}{Xin Xia},
  \bibinfo{person}{David Lo}, \bibinfo{person}{John Grundy},
  \bibinfo{person}{Xiapu Luo}, {and} \bibinfo{person}{Ting Chen}.}
  \bibinfo{year}{2020}\natexlab{}.
\newblock \showarticletitle{Defining smart contract defects on ethereum}.
\newblock \bibinfo{journal}{\emph{IEEE Transactions on Software Engineering}}
  (\bibinfo{year}{2020}).
\newblock


\bibitem[Chen et~al\mbox{.}(2021)]%
        {chen2021defectchecker}
\bibfield{author}{\bibinfo{person}{Jiachi Chen}, \bibinfo{person}{Xin Xia},
  \bibinfo{person}{David Lo}, \bibinfo{person}{John Grundy},
  \bibinfo{person}{Xiapu Luo}, {and} \bibinfo{person}{Ting Chen}.}
  \bibinfo{year}{2021}\natexlab{}.
\newblock \showarticletitle{Defectchecker: Automated smart contract defect
  detection by analyzing evm bytecode}.
\newblock \bibinfo{journal}{\emph{IEEE Transactions on Software Engineering}}
  (\bibinfo{year}{2021}).
\newblock


\bibitem[Chen et~al\mbox{.}(2019)]%
        {chen2019large}
\bibfield{author}{\bibinfo{person}{Ting Chen}, \bibinfo{person}{Zihao Li},
  \bibinfo{person}{Yufei Zhang}, \bibinfo{person}{Xiapu Luo},
  \bibinfo{person}{Ting Wang}, \bibinfo{person}{Teng Hu},
  \bibinfo{person}{Xiuzhuo Xiao}, \bibinfo{person}{Dong Wang},
  \bibinfo{person}{Jin Huang}, {and} \bibinfo{person}{Xiaosong Zhang}.}
  \bibinfo{year}{2019}\natexlab{}.
\newblock \showarticletitle{A large-scale empirical study on control flow
  identification of smart contracts}. In \bibinfo{booktitle}{\emph{2019
  ACM/IEEE International Symposium on Empirical Software Engineering and
  Measurement (ESEM)}}. IEEE, \bibinfo{pages}{1--11}.
\newblock


\bibitem[Choi et~al\mbox{.}(2021)]%
        {choi2021smartian}
\bibfield{author}{\bibinfo{person}{Jaeseung Choi}, \bibinfo{person}{Doyeon
  Kim}, \bibinfo{person}{Soomin Kim}, \bibinfo{person}{Gustavo Grieco},
  \bibinfo{person}{Alex Groce}, {and} \bibinfo{person}{Sang~Kil Cha}.}
  \bibinfo{year}{2021}\natexlab{}.
\newblock \showarticletitle{Smartian: Enhancing smart contract fuzzing with
  static and dynamic data-flow analyses}. In \bibinfo{booktitle}{\emph{2021
  36th IEEE/ACM International Conference on Automated Software Engineering
  (ASE)}}. IEEE, \bibinfo{pages}{227--239}.
\newblock


\bibitem[Christian~Reitwießner(2018)]%
        {eip165}
\bibfield{author}{\bibinfo{person}{Fabian Vogelsteller Jordi Baylina Konrad
  Feldmeier William~Entriken Christian~Reitwießner, Nick~Johnson}.}
  \bibinfo{year}{2018}\natexlab{}.
\newblock \bibinfo{title}{{EIP}-165: {Standard} {Interface} {Detection}}.
\newblock
\newblock
\urldef\tempurl%
\url{https://eips.ethereum.org/EIPS/eip-165}
\showURL{%
\tempurl}


\bibitem[Confidence interval Calculator(2023)]%
        {confidenceintervalcalculator}
Confidence interval Calculator \bibinfo{year}{2023}\natexlab{}.
\newblock \bibinfo{title}{Sample Size Calculator}.
\newblock
\newblock
\urldef\tempurl%
\url{https://www.surveysystem.com/sscalc.htm}
\showURL{%
\tempurl}


\bibitem[Das et~al\mbox{.}(2021)]%
        {das2021understanding}
\bibfield{author}{\bibinfo{person}{Dipanjan Das}, \bibinfo{person}{Priyanka
  Bose}, \bibinfo{person}{Nicola Ruaro}, \bibinfo{person}{Christopher Kruegel},
  {and} \bibinfo{person}{Giovanni Vigna}.} \bibinfo{year}{2021}\natexlab{}.
\newblock \showarticletitle{Understanding security Issues in the NFT
  Ecosystem}.
\newblock \bibinfo{journal}{\emph{arXiv preprint arXiv:2111.08893}}
  (\bibinfo{year}{2021}).
\newblock


\bibitem[Dee(2022)]%
        {dee_strategies_2022}
\bibfield{author}{\bibinfo{person}{Jim Dee}.} \bibinfo{year}{2022}\natexlab{}.
\newblock \bibinfo{title}{Strategies for {Reserved} {NFTs} in {Generative}
  {NFT} {Sets}}.
\newblock
\newblock
\urldef\tempurl%
\url{https://medium.com/web-design-web-developer-magazine/strategies-for-reserved-nfts-in-generative-nft-sets-23213db68552}
\showURL{%
\tempurl}


\bibitem[Fabian~Vogelsteller(2015)]%
        {eip20}
\bibfield{author}{\bibinfo{person}{Vitalik~Buterin Fabian~Vogelsteller}.}
  \bibinfo{year}{2015}\natexlab{}.
\newblock \bibinfo{title}{{EIP}-20: {Token} {Standard}}.
\newblock
\newblock
\urldef\tempurl%
\url{https://eips.ethereum.org/EIPS/eip-20}
\showURL{%
\tempurl}


\bibitem[Feist et~al\mbox{.}(2019)]%
        {feist2019slither}
\bibfield{author}{\bibinfo{person}{Josselin Feist}, \bibinfo{person}{Gustavo
  Grieco}, {and} \bibinfo{person}{Alex Groce}.}
  \bibinfo{year}{2019}\natexlab{}.
\newblock \showarticletitle{Slither: a static analysis framework for smart
  contracts}. In \bibinfo{booktitle}{\emph{2019 IEEE/ACM 2nd International
  Workshop on Emerging Trends in Software Engineering for Blockchain
  (WETSEB)}}. IEEE, \bibinfo{pages}{8--15}.
\newblock


\bibitem[Ferreira et~al\mbox{.}(2020)]%
        {ferreira2020smartbugs}
\bibfield{author}{\bibinfo{person}{Jo{\~a}o~F Ferreira}, \bibinfo{person}{Pedro
  Cruz}, \bibinfo{person}{Thomas Durieux}, {and} \bibinfo{person}{Rui Abreu}.}
  \bibinfo{year}{2020}\natexlab{}.
\newblock \showarticletitle{SmartBugs: a framework to analyze solidity smart
  contracts}. In \bibinfo{booktitle}{\emph{Proceedings of the 35th IEEE/ACM
  International Conference on Automated Software Engineering}}.
  \bibinfo{pages}{1349--1352}.
\newblock


\bibitem[Grieco et~al\mbox{.}(2020)]%
        {grieco2020echidna}
\bibfield{author}{\bibinfo{person}{Gustavo Grieco}, \bibinfo{person}{Will
  Song}, \bibinfo{person}{Artur Cygan}, \bibinfo{person}{Josselin Feist}, {and}
  \bibinfo{person}{Alex Groce}.} \bibinfo{year}{2020}\natexlab{}.
\newblock \showarticletitle{Echidna: effective, usable, and fast fuzzing for
  smart contracts}. In \bibinfo{booktitle}{\emph{Proceedings of the 29th ACM
  SIGSOFT International Symposium on Software Testing and Analysis}}.
  \bibinfo{pages}{557--560}.
\newblock


\bibitem[How The DAO Hack Changed Ethereum and Crypto(2023)]%
        {DAO}
How The DAO Hack Changed Ethereum and Crypto \bibinfo{year}{2023}\natexlab{}.
\newblock \bibinfo{title}{How The DAO Hack Changed Ethereum and Crypto}.
\newblock
\newblock
\urldef\tempurl%
\url{https://www.coindesk.com/consensus-magazine/2023/05/09/coindesk-turns-10-how-the-dao-hack-changed-ethereum-and-crypto/}
\showURL{%
\tempurl}


\bibitem[Jiang et~al\mbox{.}(2018)]%
        {jiang2018contractfuzzer}
\bibfield{author}{\bibinfo{person}{Bo Jiang}, \bibinfo{person}{Ye Liu}, {and}
  \bibinfo{person}{Wing~Kwong Chan}.} \bibinfo{year}{2018}\natexlab{}.
\newblock \showarticletitle{Contractfuzzer: Fuzzing smart contracts for
  vulnerability detection}. In \bibinfo{booktitle}{\emph{2018 33rd IEEE/ACM
  International Conference on Automated Software Engineering (ASE)}}. IEEE,
  \bibinfo{pages}{259--269}.
\newblock


\bibitem[Kalra et~al\mbox{.}(2018)]%
        {kalra2018zeus}
\bibfield{author}{\bibinfo{person}{Sukrit Kalra}, \bibinfo{person}{Seep Goel},
  \bibinfo{person}{Mohan Dhawan}, {and} \bibinfo{person}{Subodh Sharma}.}
  \bibinfo{year}{2018}\natexlab{}.
\newblock \showarticletitle{Zeus: analyzing safety of smart contracts.}. In
  \bibinfo{booktitle}{\emph{Ndss}}. \bibinfo{pages}{1--12}.
\newblock


\bibitem[Liu et~al\mbox{.}(2018)]%
        {liu2018reguard}
\bibfield{author}{\bibinfo{person}{Chao Liu}, \bibinfo{person}{Han Liu},
  \bibinfo{person}{Zhao Cao}, \bibinfo{person}{Zhong Chen},
  \bibinfo{person}{Bangdao Chen}, {and} \bibinfo{person}{Bill Roscoe}.}
  \bibinfo{year}{2018}\natexlab{}.
\newblock \showarticletitle{Reguard: finding reentrancy bugs in smart
  contracts}. In \bibinfo{booktitle}{\emph{2018 IEEE/ACM 40th International
  Conference on Software Engineering: Companion (ICSE-Companion)}}. IEEE,
  \bibinfo{pages}{65--68}.
\newblock


\bibitem[Luu et~al\mbox{.}(2016)]%
        {luu2016making}
\bibfield{author}{\bibinfo{person}{Loi Luu}, \bibinfo{person}{Duc-Hiep Chu},
  \bibinfo{person}{Hrishi Olickel}, \bibinfo{person}{Prateek Saxena}, {and}
  \bibinfo{person}{Aquinas Hobor}.} \bibinfo{year}{2016}\natexlab{}.
\newblock \showarticletitle{Making smart contracts smarter}. In
  \bibinfo{booktitle}{\emph{Proceedings of the 2016 ACM SIGSAC conference on
  computer and communications security}}. \bibinfo{pages}{254--269}.
\newblock


\bibitem[Lyu et~al\mbox{.}(1996)]%
        {lyu1996handbook}
\bibfield{author}{\bibinfo{person}{Michael~R Lyu} {et~al\mbox{.}}}
  \bibinfo{year}{1996}\natexlab{}.
\newblock \bibinfo{booktitle}{\emph{Handbook of software reliability
  engineering}}. Vol.~\bibinfo{volume}{222}.
\newblock \bibinfo{publisher}{IEEE computer society press Los Alamitos}.
\newblock


\bibitem[Moura and Bj{\o}rner(2008)]%
        {moura2008z3}
\bibfield{author}{\bibinfo{person}{Leonardo~de Moura} {and}
  \bibinfo{person}{Nikolaj Bj{\o}rner}.} \bibinfo{year}{2008}\natexlab{}.
\newblock \showarticletitle{Z3: An efficient SMT solver}. In
  \bibinfo{booktitle}{\emph{International conference on Tools and Algorithms
  for the Construction and Analysis of Systems}}. Springer,
  \bibinfo{pages}{337--340}.
\newblock


\bibitem[Mythril(2023)]%
        {mythril}
Mythril \bibinfo{year}{2023}\natexlab{}.
\newblock \bibinfo{title}{Mythril}.
\newblock
\newblock
\urldef\tempurl%
\url{https://mythril-classic.readthedocs.io/en/master/module-list.html}
\showURL{%
\tempurl}


\bibitem[Nguyen et~al\mbox{.}(2020)]%
        {nguyen2020sfuzz}
\bibfield{author}{\bibinfo{person}{Tai~D Nguyen}, \bibinfo{person}{Long~H
  Pham}, \bibinfo{person}{Jun Sun}, \bibinfo{person}{Yun Lin}, {and}
  \bibinfo{person}{Quang~Tran Minh}.} \bibinfo{year}{2020}\natexlab{}.
\newblock \showarticletitle{sfuzz: An efficient adaptive fuzzer for solidity
  smart contracts}. In \bibinfo{booktitle}{\emph{Proceedings of the ACM/IEEE
  42nd International Conference on Software Engineering}}.
  \bibinfo{pages}{778--788}.
\newblock


\bibitem[Nikoli{\'c} et~al\mbox{.}(2018)]%
        {nikolic2018finding}
\bibfield{author}{\bibinfo{person}{Ivica Nikoli{\'c}}, \bibinfo{person}{Aashish
  Kolluri}, \bibinfo{person}{Ilya Sergey}, \bibinfo{person}{Prateek Saxena},
  {and} \bibinfo{person}{Aquinas Hobor}.} \bibinfo{year}{2018}\natexlab{}.
\newblock \showarticletitle{Finding the greedy, prodigal, and suicidal
  contracts at scale}. In \bibinfo{booktitle}{\emph{Proceedings of the 34th
  annual computer security applications conference}}.
  \bibinfo{pages}{653--663}.
\newblock


\bibitem[OpenSea(2022)]%
        {opensea}
\bibfield{author}{\bibinfo{person}{OpenSea}.} \bibinfo{year}{2022}\natexlab{}.
\newblock \bibinfo{title}{{OpenSea}, the largest {NFT} marketplace}.
\newblock
\newblock
\urldef\tempurl%
\url{https://opensea.io/}
\showURL{%
\tempurl}


\bibitem[OpenZeppelin(2022)]%
        {erc721oz}
\bibfield{author}{\bibinfo{person}{OpenZeppelin}.}
  \bibinfo{year}{2022}\natexlab{}.
\newblock \showarticletitle{{ERC721} - {OpenZeppelin} {Docs}}.
\newblock  (\bibinfo{year}{2022}).
\newblock
\urldef\tempurl%
\url{https://docs.openzeppelin.com/contracts/4.x/erc721}
\showURL{%
\tempurl}


\bibitem[Rao et~al\mbox{.}(2012)]%
        {rao2012sailfish}
\bibfield{author}{\bibinfo{person}{Sriram Rao}, \bibinfo{person}{Raghu
  Ramakrishnan}, \bibinfo{person}{Adam Silberstein}, \bibinfo{person}{Mike
  Ovsiannikov}, {and} \bibinfo{person}{Damian Reeves}.}
  \bibinfo{year}{2012}\natexlab{}.
\newblock \showarticletitle{Sailfish: A framework for large scale data
  processing}. In \bibinfo{booktitle}{\emph{Proceedings of the Third ACM
  Symposium on Cloud Computing}}. \bibinfo{pages}{1--14}.
\newblock


\bibitem[Securify 2.0(2023)]%
        {securify2}
Securify 2.0 \bibinfo{year}{2023}\natexlab{}.
\newblock \bibinfo{title}{Securify 2.0}.
\newblock
\newblock
\urldef\tempurl%
\url{https://github.com/eth-sri/securify2}
\showURL{%
\tempurl}


\bibitem[Siakam(2022)]%
        {siakam_nft_2022}
\bibfield{author}{\bibinfo{person}{Carmen Siakam}.}
  \bibinfo{year}{2022}\natexlab{}.
\newblock \bibinfo{title}{{NFT} {MARKET}– {STATISTICS} 2021-2022}.
\newblock
\newblock
\urldef\tempurl%
\url{https://metav.rs/blog/nft-market-statistics-2021-2022/}
\showURL{%
\tempurl}


\bibitem[Smartmud(2022)]%
        {smartmud_unlimited_2022}
\bibfield{author}{\bibinfo{person}{Smartmud}.} \bibinfo{year}{2022}\natexlab{}.
\newblock \bibinfo{title}{Unlimited {Minting} of {Bored} {Ape} {Yacht} {Club}
  {NFTs}?}
\newblock
\newblock
\urldef\tempurl%
\url{www.reddit.com/r/CryptoCurrency/comments/sn8x78/unlimited_minting_of_bored_ape_yacht_club_nfts/}
\showURL{%
\tempurl}


\bibitem[Spencer(2009)]%
        {spencer2009card}
\bibfield{author}{\bibinfo{person}{Donna Spencer}.}
  \bibinfo{year}{2009}\natexlab{}.
\newblock \bibinfo{booktitle}{\emph{Card sorting: Designing usable
  categories}}.
\newblock \bibinfo{publisher}{Rosenfeld Media}.
\newblock


\bibitem[Szabo(1997)]%
        {szabo1997formalizing}
\bibfield{author}{\bibinfo{person}{Nick Szabo}.}
  \bibinfo{year}{1997}\natexlab{}.
\newblock \showarticletitle{Formalizing and securing relationships on public
  networks}.
\newblock \bibinfo{journal}{\emph{First monday}} (\bibinfo{year}{1997}).
\newblock


\bibitem[Torres et~al\mbox{.}(2018)]%
        {torres2018osiris}
\bibfield{author}{\bibinfo{person}{Christof~Ferreira Torres},
  \bibinfo{person}{Julian Sch{\"u}tte}, {and} \bibinfo{person}{Radu State}.}
  \bibinfo{year}{2018}\natexlab{}.
\newblock \showarticletitle{Osiris: Hunting for integer bugs in ethereum smart
  contracts}. In \bibinfo{booktitle}{\emph{Proceedings of the 34th Annual
  Computer Security Applications Conference}}. \bibinfo{pages}{664--676}.
\newblock


\bibitem[Tsankov et~al\mbox{.}(2018)]%
        {tsankov2018securify}
\bibfield{author}{\bibinfo{person}{Petar Tsankov}, \bibinfo{person}{Andrei
  Dan}, \bibinfo{person}{Dana Drachsler-Cohen}, \bibinfo{person}{Arthur
  Gervais}, \bibinfo{person}{Florian Buenzli}, {and} \bibinfo{person}{Martin
  Vechev}.} \bibinfo{year}{2018}\natexlab{}.
\newblock \showarticletitle{Securify: Practical security analysis of smart
  contracts}. In \bibinfo{booktitle}{\emph{Proceedings of the 2018 ACM SIGSAC
  Conference on Computer and Communications Security}}.
  \bibinfo{pages}{67--82}.
\newblock


\bibitem[Wang et~al\mbox{.}(2021)]%
        {wang2021non}
\bibfield{author}{\bibinfo{person}{Qin Wang}, \bibinfo{person}{Rujia Li},
  \bibinfo{person}{Qi Wang}, {and} \bibinfo{person}{Shiping Chen}.}
  \bibinfo{year}{2021}\natexlab{}.
\newblock \showarticletitle{Non-fungible token (NFT): Overview, evaluation,
  opportunities and challenges}.
\newblock \bibinfo{journal}{\emph{arXiv preprint arXiv:2105.07447}}
  (\bibinfo{year}{2021}).
\newblock


\bibitem[Wikipedia(2023)]%
        {confidenceinterval}
Wikipedia \bibinfo{year}{2023}\natexlab{}.
\newblock \bibinfo{title}{Confidence interval}.
\newblock
\newblock
\urldef\tempurl%
\url{https://en.wikipedia.org/wiki/Confidence\_interval}
\showURL{%
\tempurl}


\bibitem[William~Entriken(2018)]%
        {eip721}
\bibfield{author}{\bibinfo{person}{Jacob Evans Nastassia~Sachs
  William~Entriken, Dieter~Shirley}.} \bibinfo{year}{2018}\natexlab{}.
\newblock \bibinfo{title}{{EIP}-721: {Non}-{Fungible} {Token} {Standard}}.
\newblock
\newblock
\newblock
\shownote{\url{https://eips.ethereum.org/EIPS/eip-721}}.


\bibitem[Witek Radomski, Andrew Cooke, Philippe Castonguay, James Therien, Eric
  Binet, Ronan Sandford(2018)]%
        {eip1155}
Witek Radomski, Andrew Cooke, Philippe Castonguay, James Therien, Eric Binet,
  Ronan Sandford \bibinfo{year}{2018}\natexlab{}.
\newblock \bibinfo{title}{ERC-1155: Multi Token Standard}.
\newblock
\newblock
\urldef\tempurl%
\url{https://eips.ethereum.org/EIPS/eip-1155}
\showURL{%
\tempurl}


\bibitem[Zheng et~al\mbox{.}(2017)]%
        {zheng2017overview}
\bibfield{author}{\bibinfo{person}{Zibin Zheng}, \bibinfo{person}{Shaoan Xie},
  \bibinfo{person}{Hongning Dai}, \bibinfo{person}{Xiangping Chen}, {and}
  \bibinfo{person}{Huaimin Wang}.} \bibinfo{year}{2017}\natexlab{}.
\newblock \showarticletitle{An overview of blockchain technology: Architecture,
  consensus, and future trends}. In \bibinfo{booktitle}{\emph{2017 IEEE
  international congress on big data (BigData congress)}}. Ieee,
  \bibinfo{pages}{557--564}.
\newblock


\bibitem[Zheng et~al\mbox{.}(2018)]%
        {zheng2018blockchain}
\bibfield{author}{\bibinfo{person}{Zibin Zheng}, \bibinfo{person}{Shaoan Xie},
  \bibinfo{person}{Hong-Ning Dai}, \bibinfo{person}{Xiangping Chen}, {and}
  \bibinfo{person}{Huaimin Wang}.} \bibinfo{year}{2018}\natexlab{}.
\newblock \showarticletitle{Blockchain challenges and opportunities: A survey}.
\newblock \bibinfo{journal}{\emph{International journal of web and grid
  services}} \bibinfo{volume}{14}, \bibinfo{number}{4} (\bibinfo{year}{2018}),
  \bibinfo{pages}{352--375}.
\newblock


\end{thebibliography}
\bibliographystyle{ACM-Reference-Format}


\end{document}